\documentclass[12pt]{article}
\usepackage{setspace}
\usepackage[margin=1in]{geometry}
\usepackage[utf8]{inputenc}
\usepackage{amsmath}
\usepackage{amsfonts}
\usepackage{amsthm}
\usepackage{geometry}
\usepackage{multirow}
\usepackage{multicol}
\usepackage{float}
\usepackage{color,soul}
\usepackage{etoolbox}
\usepackage{titling}
\usepackage{caption}
\usepackage{subcaption}
\usepackage{graphicx}
\usepackage{booktabs}
\usepackage[english]{babel}
\usepackage [autostyle, english = american]{csquotes}

\AtBeginEnvironment{tabular}{\doublespacing}
\allowdisplaybreaks

\newtheorem{theorem}{Theorem}
\newtheorem{corollary}{Corollary}

\title{Truncation Approximation for Enriched Dirichlet Process Mixture Models}

\author{Natalie Burns$^1$ and Michael J. Daniels$^1$}

\date{%
    $^1$University of Florida\\[2ex]%
    May 2, 2023
}

\begin{document}

\begin{titlingpage}
    \maketitle
    \begin{abstract}
        Enriched Dirichlet process mixture (EDPM) models are Bayesian nonparametric models which can be used for nonparametric regression and conditional density estimation and which overcome a key disadvantage of jointly modeling the response and predictors as a Dirichlet process mixture (DPM) model: when there is a large number of predictors, the clusters induced by the DPM will be overwhelmingly determined by the predictors rather than the response. A truncation approximation to a DPM allows a blocked Gibbs sampling algorithm to be used rather than a Polya urn sampling algorithm. The blocked Gibbs sampler offers potential improvement in mixing. The truncation approximation also allows for implementation in standard software (\textit{rjags} and \textit{rstan}). In this paper we introduce an analogous truncation approximation for an EDPM. We show that with sufficiently large truncation values in the approximation of the EDP prior, a precise approximation to the EDP is available. We verify that the truncation approximation and blocked Gibbs sampler with minimum truncation values that obtain adequate error bounds achieve similar accuracy to the truncation approximation and blocked Gibbs sampler with large truncation values using a simulated example. Further, we use the simulated example to show that the blocked Gibbs sampler improves upon the mixing in the Polya urn sampler, especially as the number of covariates increases.
        
        \noindent Keywords: Bayesian nonparametrics, blocked Gibbs sampler, MCMC
    \end{abstract}
\end{titlingpage}

\section{Introduction}

Dirichlet process mixture (DPM) models are common Bayesian nonparametric (BNP) models that have a variety of applications, including density estimation \cite{escobar_bayesian_1995} and clustering \cite{do_model-based_2006}. A common use of DPMs is for nonparametric regression and conditional density estimation. Two options for achieving the latter are jointly modeling the response $Y$ and the set of predictors $X$ \cite{muller_bayesian_1996} or directly modeling the conditional distribution $Y|X=x$ \cite{maceachern_dependent_2000}. Alternatively, Wade et al's \cite{wade_improving_2014} enriched Dirichlet process mixture (EDPM) is an extension of the DPM that allows joint modelling of a response and a set of predictors while overcoming one of the disadvantages of jointly modeling $(Y,X)$ as a DPM: when there is a large number of predictors, the clusters induced by the DPM will be overwhelmingly determined by the predictors rather than the response. In this paper we introduce a truncation approximation analogous to that of Ishwaran and James for the DPM model \cite{ishwaran_approximate_2002}. We also use the truncation approximation for the EDPM to introduce a blocked Gibbs sampling algorithm for an EDPM.

\subsection{Dirichlet Process Truncation Approximation}
A DPM for data $y_1,\ldots,y_n$ with parameters $\theta_1,\ldots,\theta_n$ is given by
\begin{align*}
    y_i | \theta_i &\stackrel{ind}{\sim} F(\theta_i) \\
    \theta_i | P &\stackrel{iid}{\sim} P \\
    P &\sim DP(\alpha,P_0).
\end{align*}
Sethuraman provides a constructive definition of the DP \cite{sethuraman_constructive_1994}. In Sethuraman's stick-breaking representation, we have
\begin{align*}
    P &= \sum_{k=1}^\infty \pi_k \delta_{\theta^*_k},
\end{align*}
where $\pi_1 = V_1$; $\pi_k = V_k\prod_{h=1}^{k-1}(1-V_h)$, $k=2,3,\ldots$; $V_k \stackrel{iid}{\sim} Beta(1,\alpha)$, $k=1,2,\ldots$; $\theta_k^* \stackrel{iid}{\sim} P_{0}$, $k=1,2,\ldots$. In this representation, the atoms $\theta_k^*$ are the values of the parameters, which are shared by multiple observations, and the weights $\pi_k$ are the probabilities of drawing atom $\theta_k^*$ as the value of $\theta_i$. Ishwaran and James show that the stick-breaking representation of a DPM of normals, where $y_i|\theta_i \sim N(\mu(\theta_i),\tau(\theta_i))$, can be approximated using a finite mixture of normals:
\begin{align*}
    P_N &= \sum_{k=1}^N p_k \delta_{\theta^*_k},
\end{align*}
where $p_1 = V_1$; $p_k = V_k\prod_{h=1}^{k-1}(1-V_h)$, $k=2,3,\ldots,N$; $V_k \stackrel{iid}{\sim} Beta(1,\alpha)$, $k=1,2,\ldots,N-1$; $V_N=1$; $\theta_k^* \stackrel{iid}{\sim} P_{0}$, $k=1,2,\ldots$ \cite{ishwaran_approximate_2002}. They show that $P_N$ converges almost surely to $P$, and provide an $\mathcal{L}_1$ error bound and an error bound for the posterior of the classification variables for each observation.

The component weights, means, and variances of $P_N$ can then be estimated using a blocked Gibbs sampler in order to estimate $P_N$, which then can be used to estimate $P$ due to the truncation approximation results \cite{ishwaran_approximate_2002}. In their blocked Gibbs sampler, Ishwaran and James assume that the component distributions in the mixture are normal \cite{ishwaran_approximate_2002}. The blocked Gibbs sampling algorithm groups the means of the normal components together, the variances of the normal components together, the classification variables for all of the observations together, the weights of the normal components together, and the weights of the normal components together \cite{ishwaran_approximate_2002}. This is in contrast to the P\'olya urn Gibbs sampler in which rather than sampling the weights and atoms from the truncation approximation of the stick-breaking representation of $P$, $P$ is integrated out and the cluster assignments are sampled for each observation and the parameters are sampled for each cluster. The P\'olya urn Gibbs sampler is expected to suffer from slow mixing due to sampling observation's cluster assignments one at a time \cite{ishwaran_gibbs_2001}; the blocked Gibbs sampler does not have this issue due to the simultaneous updates of blocks of parameters \cite{ishwaran_approximate_2002}.

\subsection{Enriched Dirichlet Process}

The enriched Dirichlet process mixture (EDPM) \cite{wade_improving_2014} is an extension of the DPM that provides an alternative method for jointly modelling a response and a set of predictors. When jointly modeling $(Y,X)$ as a DPM, if there are a large number of predictors, the model may favor a large number of small clusters, and the clusters will be overwhelmingly determined by the similarity of predictors with little contribution from similarity of the response. The EDPM overcomes this disadvantage of jointly modeling $(Y,X)$ as a DPM by using a nested clustering structure with the top level clustering based on the regression of the response on the predictors and the bottom level clustering within each $Y$-cluster based on the predictors. This provides better cluster-specific estimates of the regression function and conditional density with less variability since the posterior of each $\theta^*_j$ is updated with a larger sample size based on the top level of clustering which depends only on similarity of the regression of the response on the predictors rather than also on similarity of the predictors themselves \cite{wade_improving_2014}.

Wade et al's \cite{wade_enriched_2011,wade_improving_2014} model for a response and set of covariates with an enriched Dirichlet process (EDP) prior is 
\begin{align*}
    y_i | x_i, \theta_i &\sim F_y(\cdot | x_i, \theta_i) \\
    x_i | \psi_i &\sim F_x(\cdot | \psi_i) \\
    (\theta_i, \psi_i) | P &\stackrel{iid}{\sim} P \\
    P &\sim EDP(\alpha^\theta, \alpha^{\psi|\theta}(\theta), P_0),
\end{align*}
where the EDP prior on $P$ is defined by
\begin{align*}
    P_\theta &\sim DP\left(\alpha^\theta, P_{0\theta}\right) \\
    P_{\psi | \theta}\left(\cdot|\theta\right) &\sim DP\left(\alpha^{\psi|\theta}(\theta), P_{0\psi|\theta}\left(\cdot|\theta\right)\right), \qquad\forall \theta \in \Theta,
\end{align*}
with $P_{\psi | \theta}\left(\cdot|\theta\right)$, $\forall \theta \in \Theta$, independent among themselves and of $P_\theta$, and with the mapping $\left(P_\theta, P_{\psi|\theta}\right) \rightarrow \int P_{\psi|\theta}(\cdot|\theta) dP_\theta(\theta)$. The base measure in the EDP prior is defined by $P_0(A\times B) = \int_A P_{0\psi|\theta}(B|\theta)dP_{0\theta}(\theta)$ for all Borel sets $A$ and $B$. It is not uncommon to have $P_{0\psi|\theta}$ not depend on $\theta$ so that $P_0 = P_{0\theta} \times P_{0\psi}$ \cite{wade_enriched_2011, roy_bayesian_2018}.

Wade et al \cite{wade_enriched_2011} also introduce a square-breaking construction of the EDP analogous to Sethuraman's \cite{sethuraman_constructive_1994} stick-breaking construction of the DP. If $P \sim EDP(\alpha^\theta, \alpha^{\psi|\theta}(\theta), P_0)$, then
\begin{align*}
    P = \sum_{i=1}^\infty\sum_{j=1}^\infty \pi_i^\theta\pi_{j|i}^\psi\delta_{\theta^*_i,\psi^*_{j|i}},
\end{align*}
where $\pi_1^\theta = V_1^\theta$; $\pi_i^\theta = V_i^\theta\prod_{h=1}^{i-1}(1-V_h^\theta)$, $i=2,3,\ldots$; $V_i^\theta \stackrel{iid}{\sim} Beta(1,\alpha^\theta)$, $i=1,2,\ldots$; $\theta_i^* \stackrel{iid}{\sim} P_{0\theta}$, $i=1,2,\ldots$; and for each $i=1,2,\ldots$: $\pi_{1|i}^\psi = V_{1|i}^\psi$; $\pi_{j|i}^\psi = V_{j|i}^\psi\prod_{h=1}^{j-1}(1-V_{h|i}^\psi)$, $j=2,3,\ldots$; $V_{j|i}^\psi | \theta_i^* \stackrel{iid}{\sim} Beta(1,\alpha^{\psi|\theta}(\theta_i^*))$, $j=1,2,\ldots$; and $\psi_j^* | \theta_i^* \stackrel{iid}{\sim} P_{0\psi|\theta}(\theta_i^*)$, $j=1,2,\ldots$.

The remainder of this paper is organized as follows. In section 2, we introduce a finite mixture truncation approximation for the EDP prior, show that the truncation approximation converges almost surely to an EDP, and give error bounds which show that sufficiently large numbers of components in the truncation approximation yield a precise approximation. In section 3, we introduce a blocked Gibbs sampler for an EDPM using the truncation approximation. In section 4, we use a simulated example to assess the effect of different truncation values on the accuracy of the blocked Gibbs sampler and to compare the Polya urn Gibbs sampler in \cite{wade_improving_2014} to the blocked Gibbs sampler in terms of mixing. We give conclusions and discuss open issues in section 5.

\section{EDP Truncation Approximation}

Ishwaran and James \cite{ishwaran_approximate_2002} show that we can use a finite mixture truncation approximation to the stick-breaking representation of the DP as the prior in a DPM model. The finite mixture truncation approximation prior converges almost surely to the DP, and they provide an $\mathcal{L}_1$ error bound and an error bound for the posterior of the classification variables for each observation. We extend their results to show that we can use a similar finite mixture truncation approximation for the EDP prior.

eLet
\begin{align}
    Y_i | x_i, \theta_i & \stackrel{\text{ind}}{\sim} F_y(\cdot|x_i, \theta_i) \nonumber \\
    X_i | \psi_i &\stackrel{\text{ind}}{\sim} F_x(\cdot|\psi_i) \nonumber \\
    \theta_i, \psi_i | P &\stackrel{\text{iid}}{\sim} P \nonumber \\
    P &\sim \mathcal{P}_{NM},
\end{align}
where
\begin{align*}
    \mathcal{P}_{NM}(\cdot) &= \sum_{k=1}^N\sum_{j=1}^Mp_k^{\theta}p_{j|k}^{\psi}\delta_{\theta_k^*,\psi_{j|k}^*}(\cdot),
\end{align*}
with $p_1^{\theta} = V_1^{\theta}$; $p_k^{\theta} = V_k^{\theta}\prod_{h=1}^{k-1}(1-V_h^{\theta})$, $k=2,\ldots,N$; $p_{1|k}^{\psi} = V_{1|k}^{\psi}$, $k=2,\ldots,N$; $p_{j|k}^{\psi} = V_{j|k}^{\psi}\prod_{h=1}^{j-1}(1-V_{h|k}^{\psi})$, $k=1,\ldots,N$ and $j=2,\ldots,M$; $V_k^{\theta} \stackrel{\text{iid}}{\sim} Beta(1,\alpha_{\theta})$, $k=1,\ldots,N-1,$; $V_N^{\theta}=1$; and for $k=1,\ldots,N$, $V_{j|k}^{\psi}|\theta_k^* \stackrel{\text{iid}}{\sim} Beta(1,\alpha^{\psi|\theta}(\theta_k^*))$, $j=1,\ldots,M-1,$ and $V_{M|k}^{\psi}=1.$ Also, $\theta_k^* \stackrel{\text{iid}}{\sim} P_{0\theta}$ for $k=1,\ldots,N$ and $\psi_{j|k}^* | \theta_k^* \stackrel{\text{iid}}{\sim} P_{0\psi|\theta}(\cdot|\theta_k^*)$ for $k=1,\ldots,N$ and $j=1,\ldots,M$.

First, note that $\mathcal{P}_{NM}$ converges almost surely to an enriched Dirichlet process with base distribution $P_{0\theta}\times P_{0\psi|\theta}$ and precision parameters $\alpha^\theta, \alpha^{\psi|\theta}$.  Let
\begin{align*}
    \mathcal{P}_{\infty}(\cdot) = EDP(\alpha^\theta, \alpha^{\psi|\theta}, P_0) = \sum_{k=1}^\infty\sum_{j=1}^\infty \pi_k^{\theta}\pi_{j|k}^{\psi}\delta_{\theta_k^*,\psi_{j|k}^*}(\cdot),
\end{align*}
where $\pi_1^{\theta} = Z_1^{\theta}$; $\pi_k^{\theta} = Z_k^{\theta}\prod_{h=1}^{k-1}(1-Z_h^{\theta})$, $k=2,3,\ldots$; $\pi_{1|k}^{\psi} = Z_{1|k}^{\psi}$, $k=1,2,\ldots$; $\pi_{j|k}^{\psi} = Z_{j|k}^{\psi}\prod_{h=1}^{j-1}(1-Z_{h|k}^{\psi})$, $k=1,2,\ldots$ and $j=2,3,\ldots$; $V_k^{\theta} \stackrel{\text{iid}}{\sim} Beta(1,\alpha_{\theta})$, $k=2,3,\ldots$; $\theta_k^* \stackrel{\text{iid}}{\sim} P_{0\theta}$, $k=1,2,\ldots$; and for $k=1,2,\ldots$, $Z_{j|k}^{\psi}|\theta_k^* \stackrel{\text{iid}}{\sim} Beta(1,\alpha^{\psi|\theta}(\theta_k^*))$ and $\psi_{j|k}^* | \theta_k^* \stackrel{\text{iid}}{\sim} P_{0\psi|\theta}(\cdot|\theta_k^*)$ for $j=1,2,\ldots$.  Letting $D(\mathbb{P}_1,\mathbb{P}_2)=\sup_A|\mathbb{P}_1(A)-\mathbb{P}_2(A)|$ be the total variation distance between two probability measures, $\mathbb{P}_1$ and $\mathbb{P}_2$, we have
\begin{align*}
    \Pr\left(\lim_{N,M \to \infty} D(\mathcal{P}_{NM},\mathcal{P}_\infty) = 0\right) &= \Pr\left(\lim_{N,M \to \infty} \sup_A |\mathcal{P}_\infty(A) - \mathcal{P}_{NM}(A)| = 0\right) \\
    &= \Pr\left(\lim_{N,M \to \infty} \sup_A \left| \sum_{k=1}^\infty\sum_{j=1}^\infty \pi_k^{\theta}\pi_{j|k}^{\psi}\delta_{\theta_k^*,\psi_{j|k}^*}(A) \right.\right. \\
    &\phantom{= \Pr(\lim_{N,M \to \infty}\sup_A |\sum_{k=1}^\infty}\left. \left. - \sum_{k=1}^N\sum_{j=1}^M p_k^{\theta}p_{j|k}^{\psi}\delta_{\theta_k^*,\psi_{j|k}^*}(A)\right| = 0\right),
\end{align*}
and since the values of $\pi_k^{\theta}$ and $p_k^{\theta}$, $\pi_{j|k}^{\psi}$ and $p_{j|k}^{\psi}$, $\theta_k^*$, and $\psi_{j|k}^*$ will be the same under $\mathcal{P}_\infty$ and $\mathcal{P}_{NM}$, respectively, within the $k$th $\theta$-cluster and $j$th $\psi$-cluster when $k<N$ and $j<M$, with probability 1 we have
\begin{align*}
    D(\mathcal{P}_{NM},\mathcal{P}_\infty) &= \sup_A \left| \sum_{k=N}^\infty\sum_{j=1}^\infty \pi_k^{\theta}\pi_{j|k}^{\psi}\delta_{\theta_k^*,\psi_{j|k}^*}(A)\right. - \sum_{j=1}^M p_N^{\theta}p_{j|N}^{\psi}\delta_{\theta_N^*,\psi_{j|N}^*}(A) \\
    &\phantom{=\Pr\Bigg(} \left. + \sum_{k=1}^{N-1}\left[\sum_{j=M}^\infty\pi_k^{\theta}\pi_{j|k}^{\psi}\delta_{\theta_k^*,\psi_{j|k}^*}(A) - p_k^{\theta}p_{M|k}^{\psi}\delta_{\theta_k^*,\psi_{M|k}^*}(A)\right] \right| \\
    &\leq \sup_A \left[\sum_{k=N}^\infty\sum_{j=1}^\infty \pi_k^{\theta}\pi_{j|k}^{\psi}\delta_{\theta_k^*,\psi_{j|k}^*}(A)\right] + \sup_A \left[\sum_{j=1}^M p_N^{\theta}p_{j|N}^{\psi}\delta_{\theta_N^*,\psi_{j|N}^*}(A)\right] \\
    &\phantom{=\Pr\Bigg(} + \sum_{k=1}^{N-1}\left\{\sup_A \left[\sum_{j=M}^\infty\pi_k^{\theta}\pi_{j|k}^{\psi}\delta_{\theta_k^*,\psi_{j|k}^*}(A)\right] + \sup_A\left[p_k^{\theta}p_{M|k}^{\psi}\delta_{\theta_k^*,\psi_{M|k}^*}(A)\right]\right\} \\
    &= \sum_{k=N}^\infty\pi_k^{\theta}\sum_{j=1}^\infty \pi_{j|k}^{\psi} + p_N^{\theta}\sum_{j=1}^M p_{j|N}^{\psi} + \sum_{k=1}^{N-1}\left\{ \pi_k^{\theta}\sum_{j=M}^\infty\pi_{j|k}^{\psi} + p_k^{\theta}p_{M|k}^{\psi}\right\} \\
    &= \left(1 - \sum_{k=1}^{N-1}\pi_k^{\theta}\right) + \left(1 - \sum_{k=1}^{N-1}\pi_k^{\theta}\right)\sum_{j=1}^M p_{j|N}^{\psi} \\
    &\phantom{=\Pr\Bigg(} + \sum_{k=1}^{N-1}\left\{ \pi_k^{\theta}\left(1-\sum_{j=1}^{M-1}\pi_{j|k}^{\psi}\right) + p_k^{\theta}\left(1-\sum_{j=1}^{M-1}\pi_{j|k}^{\psi}\right)\right\} \\
    &\xrightarrow[M \to \infty]{} 2\left(1 - \sum_{k=1}^{N-1}\pi_k^{\theta}\right) \\
    &\xrightarrow[N \to \infty]{} 0.
\end{align*}
Thus, $\mathcal{P}_{NM}$ converges almost surely to the EDP.

Sufficiently large values of $N$ and $M$ will give a precise approximation by yielding a marginal density for $(\mathbf{y},\mathbf{x})$ which is close to the marginal density when we use the EDP instead of the truncation approximation.  Let
\begin{align*}
    m_{NM}(\mathbf{y},\mathbf{x}) &= \int\left[\prod_{i=1}^n\int f_y(y_i|x_i, \theta_i)f_x(x_i|\psi_i)dP(\theta_i,\psi_i)\right]d\mathcal{P}_{NM}(P),
\end{align*}
be the marginal density under (1), where $f_y(y_i|x_i,\theta_i)$ and $f_x(x_i|\psi_i)$ are the densities associated with $F_y(\cdot|x,\theta)$ and $F_x(\cdot|\psi)$, respectively, and let $m_\infty(\mathbf{y},\mathbf{x})$ be the marginal density under (1) but with $P\sim EDP(\alpha^\theta, \alpha^{\psi|\theta}, P_0)$.  In the case where $\alpha^{\psi|\theta}$ does not depend on $\theta^*_{k}$ we have the following $\mathcal{L}_1$ error bound. The proof can be found in Appendix A.

\begin{theorem} \label{thm1}
We have
\begin{align}
    \int_{\mathbb{R}^n\times\mathbb{R}^n} |m_{NM}(\mathbf{y},\mathbf{x}) - m_\infty(\mathbf{y},\mathbf{x})|d(\mathbf{y},\mathbf{x}) &\leq 4\left[1 - E\left\{\left(\sum_{k=1}^{N-1}p_k^{\theta}\sum_{j=1}^{M-1}p_{j|k}^{\psi}\right)^n\right\}\right] \nonumber \\
    &\approx 4n\left[\exp{\left\{-\frac{N-1}{\alpha^\theta}\right\}} \right. \nonumber \\
    &\phantom{\approx 4n[\exp} \left. + \exp{\left\{-\frac{M-1}{\alpha^\psi}\right\}}\left(1 - \exp{\left\{-\frac{N-1}{\alpha^\theta}\right\}}\right)\right],
\end{align}
where the $p_k^{\theta}$ and $p_{j|k}^{\psi}$ are defined as above.
\end{theorem}

It is a simple adjustment to allow for the case where the value of $\alpha^{\psi|\theta}$ does depend on with which $\theta$-cluster it is associated. Let $\alpha^{\psi|\theta}_k = \alpha^{\psi|\theta}(\theta^*_k)$ for each $k = 1, 2, \ldots$; that is, $\alpha^{\psi|\theta}_k$ is the value of $\alpha^{\psi|\theta}$ associated with the $k$th $\theta$-cluster but does not otherwise depend on the value of $\theta_k^*$. The $\mathcal{L}_1$ error bound for this case is given in Corollary 1 and details can be found in Appendix A.

\begin{corollary} \label{cor1}
We have
\begin{align}
    \int_{\mathbb{R}^n\times\mathbb{R}^n} |m_{NM}(\mathbf{y},\mathbf{x}) - m_\infty(\mathbf{y},\mathbf{x})|d(\mathbf{y},\mathbf{x}) &\leq 4\left[1 - E\left\{\left(\sum_{k=1}^{N-1}p_k^{\theta}\sum_{j=1}^{M-1}p_{j|k}^{\psi}\right)^n\right\}\right] \nonumber \\
    &\approx 4n\left(\exp{\left\{-\frac{N-1}{\alpha^\theta}\right\}} + E\left[\sum_{k=1}^{N-1}p_k^{\theta}\exp{\left\{-\frac{M-1}{\alpha^{\psi|\theta}_k}\right\}}\right]\right) \nonumber \\
    &\leq 4n\left(\exp{\left\{-\frac{N-1}{\alpha^\theta}\right\}} + \exp{\left\{-\frac{M-1}{\alpha^{\psi|\theta}_\text{max}}\right\}}E\left[\sum_{k=1}^{N-1}p_k^{\theta}\right]\right) \nonumber \\
    &\approx 4n\left[\exp{\left\{-\frac{N-1}{\alpha^\theta}\right\}} \right. \nonumber \\
    &\phantom{\approx 4n[\exp} \left. + \exp{\left\{-\frac{M-1}{\alpha^{\psi|\theta}_\text{max}}\right\}}\left(1 - \exp{\left\{-\frac{N-1}{\alpha^\theta}\right\}}\right)\right],
\end{align}
where the $p_k^{\theta}$ and $p_{j|k}^{\psi}$ are defined as above, and $\alpha^{\psi|\theta}_\text{max}$ is the largest of $\alpha^{\psi|\theta}_1,\ldots,\alpha^{\psi|\theta}_N$.
\end{corollary}

We can empirically assess the effect of varying values of $N$ and $M$ on the $\mathcal{L}_1$ error bounds given in Theorem 1 and Corollary 1 by considering different sample sizes and values of the concentration parameters in the EDP. To illustrate, we consider sample sizes $n=200$ and $n=1000$; smaller values of each concentration parameter $\alpha^{\theta} = 0.5$ or $\alpha^{\psi} = 0.5$, which favor smaller numbers of $\theta$- and $\psi$-clusters, respectively; and larger values of each concentration parameter $\alpha^{\theta} = 3$ or $\alpha^{\psi} = 3$, which favor larger numbers of $\theta$- and $\psi$-clusters, respectively.

Considering a smaller sample size of $n=200$, if we have relatively small values of both concentration parameters, $\alpha^\theta=0.5$ and $\alpha^\psi=0.5$, as few as 10 clusters for the truncation values of $N$ and $M$ yield a very close approximation with an $\mathcal{L}_1$ error bound of $2.437\times 10^{-5}$. If one of the concentration parameters takes a large value, $\alpha^\psi=3$ for example, increasing the corresponding number of clusters to $M=50$ gives an $\mathcal{L}_1$ error bound of $7.669\times 10^{-5}$. Finally, if both concentration parameters take a large value with $\alpha^\theta=3$ and $\alpha^\psi=3$, using 50 clusters for the truncation values of both $N$ and $M$ yield a very close approximation with an $\mathcal{L}_1$ error bound of $1.290\times 10^{-4}$.

Similarly, considering a larger sample size of $n=1000$, even for large values of the concentration parameters $\alpha^\theta=3$ and $\alpha^\psi=3$, 50 clusters for the truncation values of $N$ and $M$ yield an $\mathcal{L}_1$ error bound of $6.451\times 10^{-4}$. In general, to obtain an adequate approximation to the untruncated EDP, larger truncation values are necessary for larger sample sizes, larger values of $N$ are necessary for larger values of $\alpha^\theta$, and larger values of $M$ are necessary for larger values of $\alpha^\psi$. These general principles apply regardless of whether we are using a fixed $\alpha^\psi$ for all $\theta$-clusters, or if we are allowing $\alpha^{\psi|\theta}_k$ to vary depending on which $\theta$-cluster it is associated with and using $\alpha^{\psi|\theta}_{max}$ to calculate the error bounds. Table \ref{tab:NM} illustrates the effects of different sample sizes and values of the concentration parameters $\alpha^\theta$ and $\alpha^\psi$ on the truncation values necessary for an adequate approximation to the untruncated EDP. Note that in general, the values of the concentration parameters are unknown, but we can obtain estimates of $\alpha^\theta$ and either $\alpha^\psi$ or $\alpha^{
\psi|\theta}_{max}$ using a short run of the blocked Gibbs sampler using large values of $N$ and $M$.

\begin{table}
    \centering
    \begin{tabular}{|c|c|c|c|c|}
         \hline
         $\alpha^\theta$ & $\alpha^\psi$ & $n=200$ & $n=1,000$ & $n=2,000$ \\
         \hline
         0.5 & 0.5 & $N=7$, $M=7$ & $N=8$, $M=8$ & $N=8$, $M=9$ \\
         \hline
         0.5 & 1.5 & $N=7$, $M=19$ & $N=8$, $M=21$ & $N=8$, $M=24$ \\
         \hline
         0.5 & 3 & $N=7$, $M=37$ & $N=8$, $M=41$ & $N=8$, $M=46$ \\
         \hline
         1.5 & 1.5 & $N=19$, $M=19$ & $N=21$, $M=21$ & $N=23$, $M=23$ \\
         \hline
         3 & 0.5 & $N=37$, $M=7$ & $N=41$, $M=8$ & $N=44$, $M=9$ \\
         \hline
         3 & 3 & $N=37$, $M=37$ & $N=41$, $M=41$ & $N=44$, $M=44$  \\
         \hline
    \end{tabular}
    \caption{Minimum truncation values $N$ and $M$ which give an $\mathcal{L}_1$ error bound less than 0.01 for varying sample sizes and values of the concentration parameters $\alpha^\theta$ and $\alpha^\psi$.}
    \label{tab:NM}
\end{table}

Using the $\mathcal{L}_1$ error bound from Theorem 1, we can also obtain an error bound for the posterior of the classification variables $K_1,\ldots,K_n$ and $J_1,\ldots,J_n$, where $K_i$ gives the $\theta$-cluster and $J_i$ gives the $\psi$-cluster within the $\theta$-cluster of the $i$th observation.  The error bound is given in the following corollary. The proof can again be found in Appendix A.

\begin{corollary} \label{cor2}
We have
\begin{align*}
    \int_{\mathbb{R}^n\times\mathbb{R}^n}\Bigg(\sum_{\mathbf{K}\in\mathcal{K}_\infty}&\sum_{\mathbf{J}\in\mathcal{J}_\infty} |\pi_{NM}(\mathbf{K},\mathbf{J}|\mathbf{y},\mathbf{x}) - \pi_{\infty}(\mathbf{K},\mathbf{J}|\mathbf{y},\mathbf{x})|\Bigg) m_\infty(\mathbf{y},\mathbf{x})d(\mathbf{y},\mathbf{x}) \\
    &= O\left(n\left(\exp{\left\{-\frac{N-1}{\alpha^\theta}\right\}} + \exp{\left\{-\frac{M-1}{\alpha^\psi}\right\}}\left(1 - \exp{\left\{-\frac{N-1}{\alpha^\theta}\right\}}\right)\right)\right),
\end{align*}
where $\pi_{NM}(\mathbf{K},\mathbf{J}|\mathbf{y},\mathbf{x})$ and $\pi_{\infty}(\mathbf{K},\mathbf{J}|\mathbf{y},\mathbf{x})$ are the posteriors of $(\mathbf{K},\mathbf{J})$ under $\mathcal{P}_{NM}$ and $EDP(\alpha^\theta, \alpha^\psi, P_0)$, respectively.
\end{corollary}

Again a simple adjustment alllows for the case where the value of $\alpha^{\psi|\theta}$ depends on with which $\theta$-cluster it is associated. The error bound for the posterior of the classification variables for this case is given in Corollary 3 and details can be found in Appendix A.

\begin{corollary} \label{cor3}
We have 
\begin{align*}
    \int_{\mathbb{R}^n\times\mathbb{R}^n}\Bigg(\sum_{\mathbf{K}\in\mathcal{K}_\infty}&\sum_{\mathbf{J}\in\mathcal{J}_\infty} |\pi_{NM}(\mathbf{K},\mathbf{J}|\mathbf{y},\mathbf{x}) - \pi_{\infty}(\mathbf{K},\mathbf{J}|\mathbf{y},\mathbf{x})|\Bigg) m_\infty(\mathbf{y},\mathbf{x})d(\mathbf{y},\mathbf{x}) \\
    &= O\left(n\left(\exp{\left\{-\frac{N-1}{\alpha^\theta}\right\}} + \exp{\left\{-\frac{M-1}{\alpha^{\psi|\theta}_\text{max}}\right\}}\left(1 - \exp{\left\{-\frac{N-1}{\alpha^\theta}\right\}}\right)\right)\right),
\end{align*}
where $\pi_{NM}(\mathbf{K},\mathbf{J}|\mathbf{y},\mathbf{x})$ and $\pi_{\infty}(\mathbf{K},\mathbf{J}|\mathbf{y},\mathbf{x})$ are the posteriors of $(\mathbf{K},\mathbf{J})$ under $\mathcal{P}_{NM}$ and $EDP(\alpha^\theta, \alpha^{\psi|\theta}, P_0)$, respectively, and $\alpha^{\psi|\theta}_\text{max}$ is the largest of $\alpha^{\psi|\theta}_1,\ldots,\alpha^{\psi|\theta}_N$.
\end{corollary}

\section{Blocked Gibbs Sampler for the EDP Truncation Approximation}

Similar to Ishwaran and James' blocked Gibbs sampler for the DPM of normals, we develop a blocked Gibbs sampler for an EDPM of normals using the truncation approximation introduced in Section 2. We have $Y_i|x_i,\theta_i \stackrel{\text{ind}}{\sim} N(\mathbf{x}_i\boldsymbol{\beta}_i,\tau_{yi})$, $\theta_i=(\boldsymbol{\beta}_{i},\tau_{yi})$, and $X_{il}|\psi_i \stackrel{\text{ind}}{\sim} N(\boldsymbol{\mu}_{il},\tau_{x,il})$ for each covariate $X_{il}$, $l=1,\ldots,p$, $\psi_i=(\mu_{xi1},\tau_{xi1},\ldots,\mu_{xip},\tau_{xip})$. The model for this EDPM is as follows:
\begin{align*}
    y_i | \mathbf{x}_i,  \boldsymbol{\theta}_i &\sim N\left(\mathbf{x}^*_i\boldsymbol{\beta}_{i}, \tau_{y,i}\right) \\
    x_{i,l} | \boldsymbol{\psi}_i &\sim N\left(\mu_{il}, \tau_{x,il}\right) \\
    \left(\boldsymbol{\theta}_i, \boldsymbol{\psi}_i\right) | P &\sim P \\
    P &\sim EDP\left(\alpha_\theta, \alpha_\psi(\theta), P_0\right),
\end{align*}
where $\mathbf{x}^*_i$ is the vector of covariates for the $i$th subject including an intercept. We use the base measure $P_0 = P_{0\theta} \times P_{0\psi|\theta}$, where
\begin{align*}
    P_{0\theta} &\sim N\left(\boldsymbol{\beta}; \boldsymbol{\beta}_{0}, \tau_{y} \mathbf{C}_y^{-1}\right) \times IG\left(\tau_{y}; a_y, b_y\right) \\
    P_{0\psi|\theta} &\sim \prod_{l=1}^{p}N\left(\mu_l; m_{l}, \tau_{x,l} c_{x,l}^{-1}\right) \times IG\left(\tau_{x,l}; a_{x}, b_{x}\right).
\end{align*}
Note that we could also include binary or categorical covariates with their corresponding conjugate priors included in $P_{0\psi|\theta}$.

Recall that observations from the same $\theta$-cluster will share values of $\theta$ and observations from the same $\psi$-cluster within the same $\theta$-cluster will share values of $\psi$. We denote the $N\times(p+1)$ matrix of $\theta$-cluster coefficients as $\boldsymbol{\beta}^*$, the vector of $\theta$-cluster variances as $\boldsymbol{\tau}^*_y$, and the vectors of $\psi$-cluster means and variances for the $l$th covariate within the $k$th $\theta$-cluster as $\boldsymbol{\mu}^*_{kl}$ and $\boldsymbol{\tau}^*_{xkl}$, respectively. In the blocked Gibbs sampler, we sample from the following conditional distributions iteratively:
\begin{align*}
    &(\boldsymbol{\beta}^* | \boldsymbol{\tau}^*_y, \mathbf{K}, \mathbf{Y}) \\
    &(\boldsymbol{\tau}^*_y | \boldsymbol{\mu}^*_y, \mathbf{K}, \mathbf{Y}) \\
    &(\boldsymbol{\mu}^*_{kl} | \boldsymbol{\tau}^*_{xkl}, \mathbf{K}, \mathbf{J}, \mathbf{X}), \qquad k=1,\ldots,N \text{ and } l=1,\ldots,p \\
    &(\boldsymbol{\tau}^*_{xkl} | \boldsymbol{\mu}^*_{kl}, \mathbf{K}, \mathbf{J}, \mathbf{X}), \qquad k=1,\ldots,N \text{ and } l=1,\ldots,p \\
    &(\mathbf{K}, \mathbf{J} | \mathbf{p}^\theta, \mathbf{p}^\psi, \boldsymbol{\beta}^*, \boldsymbol{\tau}^*_y, \boldsymbol{\mu}^*_{xk}, \boldsymbol{\tau}^*_{xk}, \mathbf{Y}, \mathbf{X}) \\
    &(\mathbf{p}^\theta | \mathbf{K}, \alpha^\theta) \\
    &(\mathbf{p}^\psi_k | \mathbf{K}, \mathbf{J}, \alpha^{\psi|\theta}_k), \qquad k=1,\ldots,N \\
    &(\alpha^\theta|\mathbf{p}^\theta) \\
    &(\alpha^{\psi|\theta}_k|\mathbf{p}^\psi), \qquad k=1,\ldots,N.
\end{align*}
Each of these updates is available in closed form; details can be found in Appendix B. At each iteration $b$ of the sampler, we can then use the $\theta$-component weights, regression coefficients, and variances and the $\psi$-component weights, means, and variances to obtain a draw from the posterior $\mathcal{P}_{NM}|\mathbf{Y},\mathbf{X}$:
\begin{align*}
    \mathcal{P}_{NM}^b(\cdot) &= \sum_{k=1}^N\sum_{j=1}^Mp_k^{\theta, b}p_{j|k}^{\psi, b}\delta_{\theta_k^{* b},\psi_{j|k}^{* b}}(\cdot).
\end{align*}

The blocked Gibbs sampler can also be adapted to non-normal mixtures, and is particularly straightforward when conjugate updates are available. The more general model for the EDPM is as follows:
\begin{align*}
    y_i | \mathbf{x}_i,  \boldsymbol{\theta}_i &\sim F_y\left(\mathbf{x}_i; \boldsymbol{\theta}_{i}\right) \\
    x_{i,l} | \boldsymbol{\psi}_i &\sim F_x\left(\psi_{il}\right) \\
    \left(\boldsymbol{\theta}_i, \boldsymbol{\psi}_i\right) | P &\sim P \\
    P &\sim EDP\left(\alpha_\theta, \alpha_\psi(\theta), P_0\right),
\end{align*}
and in the blocked Gibbs sampler, we sample from the following conditional distributions iteratively:
\begin{align*}
    &(\boldsymbol{\theta}^* | \mathbf{K}, \mathbf{Y}) \\
    &(\boldsymbol{\psi}^*_{kl} | \mathbf{K}, \mathbf{J}, \mathbf{X}), \qquad k=1,\ldots,N \text{ and } l=1,\ldots,p \\
    &(\mathbf{K}, \mathbf{J} | \mathbf{p}^\theta, \mathbf{p}^\psi, \boldsymbol{\theta}^*, \boldsymbol{\psi}^*_{xk}, \mathbf{Y}, \mathbf{X}) \\
    &(\mathbf{p}^\theta | \mathbf{K}, \alpha^\theta) \\
    &(\mathbf{p}^\psi_k | \mathbf{K}, \mathbf{J}, \alpha^{\psi|\theta}_k), \qquad k=1,\ldots,N \\
    &(\alpha^\theta|\mathbf{p}^\theta) \\
    &(\alpha^{\psi|\theta}_k|\mathbf{p}^\psi), \qquad k=1,\ldots,N.
\end{align*}

\section{Simulations}

Wade et al \cite{wade_improving_2014} used a toy example to demonstrate the advantages of the EDP over jointly modeling a response and predictors using a DPM. We now consider the same toy example to assess the accuracy of the truncation approximation with different truncation values and to compare the performance of the blocked Gibbs sampler to the performance of Wade et al's algorithm (a Polya urn sampler) based on Algorithm 8 of Neal \cite{neal_markov_2000}. In the example in \cite{wade_improving_2014}, the true model of the response $Y$ is mixture of two normals which depend upon only the first covariate:
\begin{align*}
    Y_i | x_i &\stackrel{ind}{\sim} p(x_{i,1})N(y_i|\beta_{1,0} + \beta_{1,1}x_{i,1}, \sigma_1^2) + (1-p(x_{i,1}))N(y_i|\beta_{2,0} + \beta_{2,1}x_{i,1}, \sigma_2^2),
\end{align*}
where
\begin{align*}
    p(x_{i,1}) &= \frac{\omega_1\exp{\left(\frac{-\omega_1}{2}(x_{i,1}-\mu_1)^2\right)}}{\omega_1\exp{\left(\frac{-\omega_1}{2}(x_{i,1}-\mu_1)^2\right)} + \omega_2\exp{\left(\frac{-\omega_2}{2}(x_{i,1}-\mu_2)^2\right)}},
\end{align*}
and $\beta_1=(0,1)^T$, $\sigma_1^2=1/16$, $\beta_2=(4.5,0.1)^T$, $\sigma_2^2=1/8$, $\mu_1=4$, $\mu_2=6$, and $\omega_1=\omega_2=2$. The covariates are drawn from a multivariate normal distribution,
\begin{align*}
    \mathbf{X}_i &= (X_{i,1},\ldots,X_{i,p}) \stackrel{iid}{\sim} N(\mu,\Sigma), 
\end{align*}
with mean $\mu = (4,\ldots,4)$ and variance-covariance matrix $\Sigma$ where $\Sigma_{h,h} = 4$, $\Sigma_{h,l}=3.5$ for $h\neq l$ in $\{1,2,4,\ldots,2\lfloor p/2\rfloor\}$ or $h\neq l$ in $\{3,5,\ldots,2\lfloor (p-1)/2\rfloor+1\}$, and $\Sigma_{h,l}=0$ for all other $h\neq l$. For each of $p=5$, 10, and 15, we draw 10 samples of size $n=200$ from this model. As noted in Section 3, we use the conjugate base measure. Finally, we place a $Gamma(1,1)$ prior on $\alpha^\theta$ and each $\alpha^{\psi|\theta}_k$.

\subsection{Accuracy}

We assess the accuracy of the truncation approximation by running the blocked Gibbs sampler both with the minimum truncation values that provide adequately small error bounds and with larger truncation values of $N=10$ and $M=50$ for each of the samples. We obtain posterior inference with 100,000 iterations with a burn in period of 20,000 iterations. To determine the truncation values to use in the blocked Gibbs sampler with smaller truncation values, after completing a short run of 10,000 iterations of the blocked Gibbs sampler for each of the samples, we estimated $\alpha^\theta$ and $\alpha_{max}^{\psi|\theta}$ in order to calculate the $\mathcal{L}_1$ error bound for various truncation values for each of the samples, as described in Section 2. We then use the smallest possible truncation values that obtain error bounds less than or equal to 0.01. For example, for one set of samples, one each with $p=5$, 10, and 15, the truncation values $N=10$ and $M=50$ give error bounds of $2.105 \times 10^{-5}$, $6.251 \times 10^{-4}$, and $3.248 \times 10^{-10}$, respectively. In order to obtain error bounds less than or equal to 0.01, for these samples we use truncation values $N=7$ and $M=33$ for $p=5$, $N=9$ and $M=23$ for $p=10$, and $N=7$ and $M=21$ for $p=15$.

We first assess the accuracy of the blocked Gibbs samplers using one simulated data for each  $p \in \{5,10,15\}$ and computing $E[Y|X=x]$ for 20 new subjects with values of $x_1$ evenly spaced between -0.5 and 8. Figure 1 shows the results for the blocked Gibbs (BG) sampler with fixed truncation values of $N=10$ and $M=50$ and the blocked Gibbs (BG) sampler with varying truncation values depending on $p$ that obtain sufficiently small error bounds as described above. We can see that the estimates using the blocked Gibbs sampler with either set of truncation values are similarly close. The width of the 95\% credible intervals using the blocked Gibbs sampler with either set of truncation values are similar as well.

\begin{figure}
    \centering
    \begin{subfigure}[h]{0.32\textwidth}
        \centering
        \includegraphics[width=\linewidth]{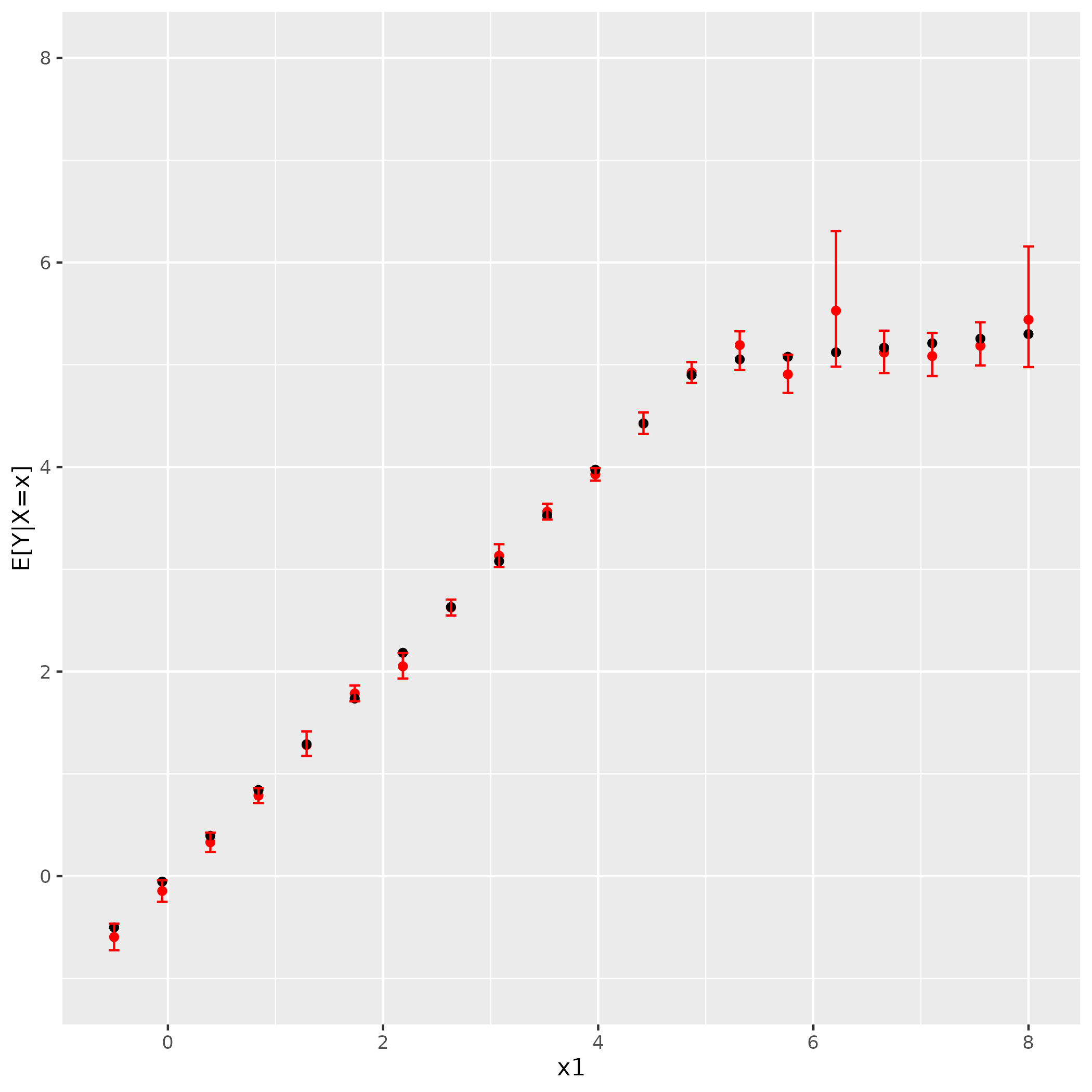} 
        \caption{BG, $p=5$, $N=10$, $M=50$} \label{fig:BG5}
    \end{subfigure}
    \hfill
    \begin{subfigure}[h]{0.33\textwidth}
        \centering
        \includegraphics[width=\linewidth]{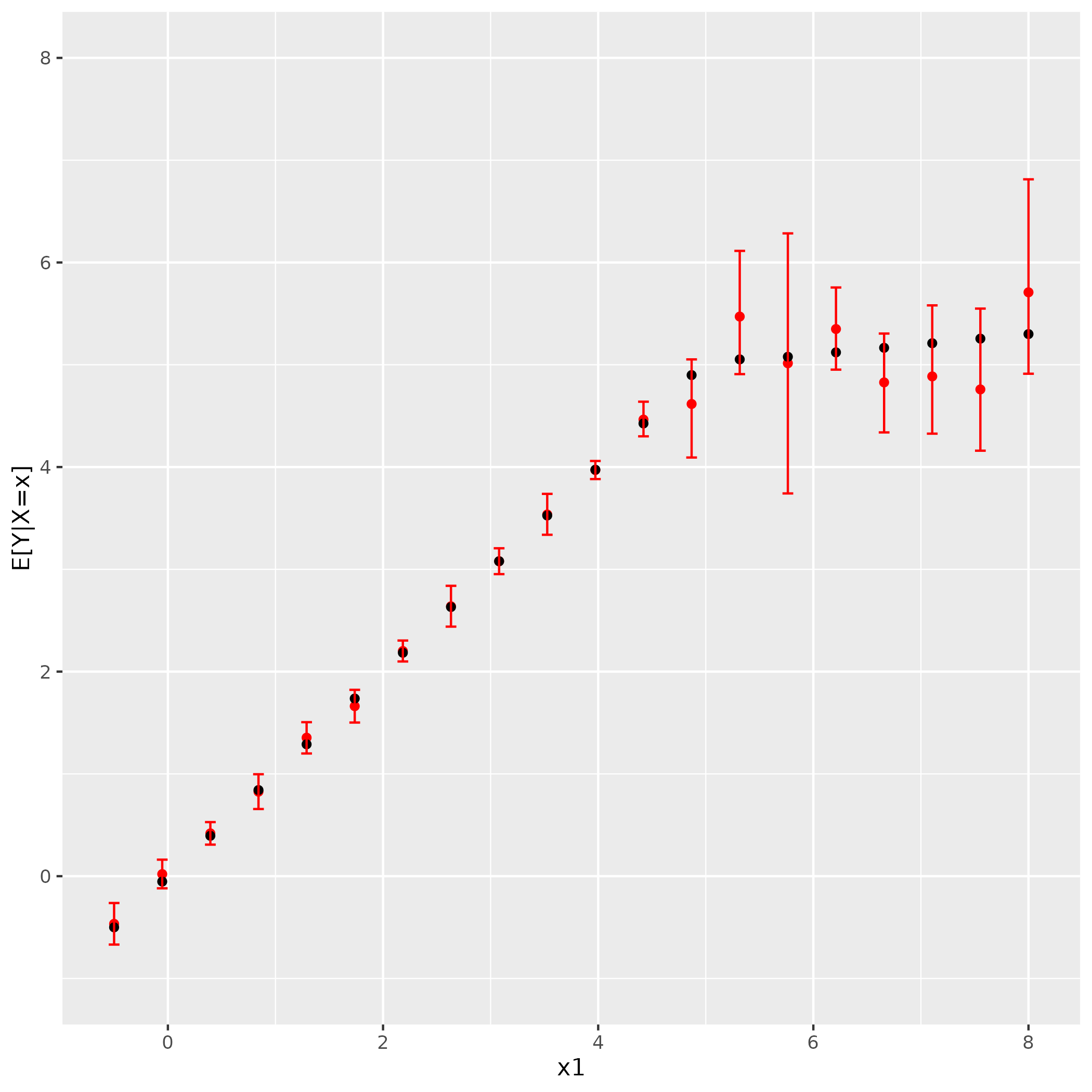} 
        \caption{BG, $p=10$, $N=10$, $M=50$} \label{fig:BG10}
    \end{subfigure}
    \hfill
    \begin{subfigure}[h]{0.33\textwidth}
        \centering
        \includegraphics[width=\linewidth]{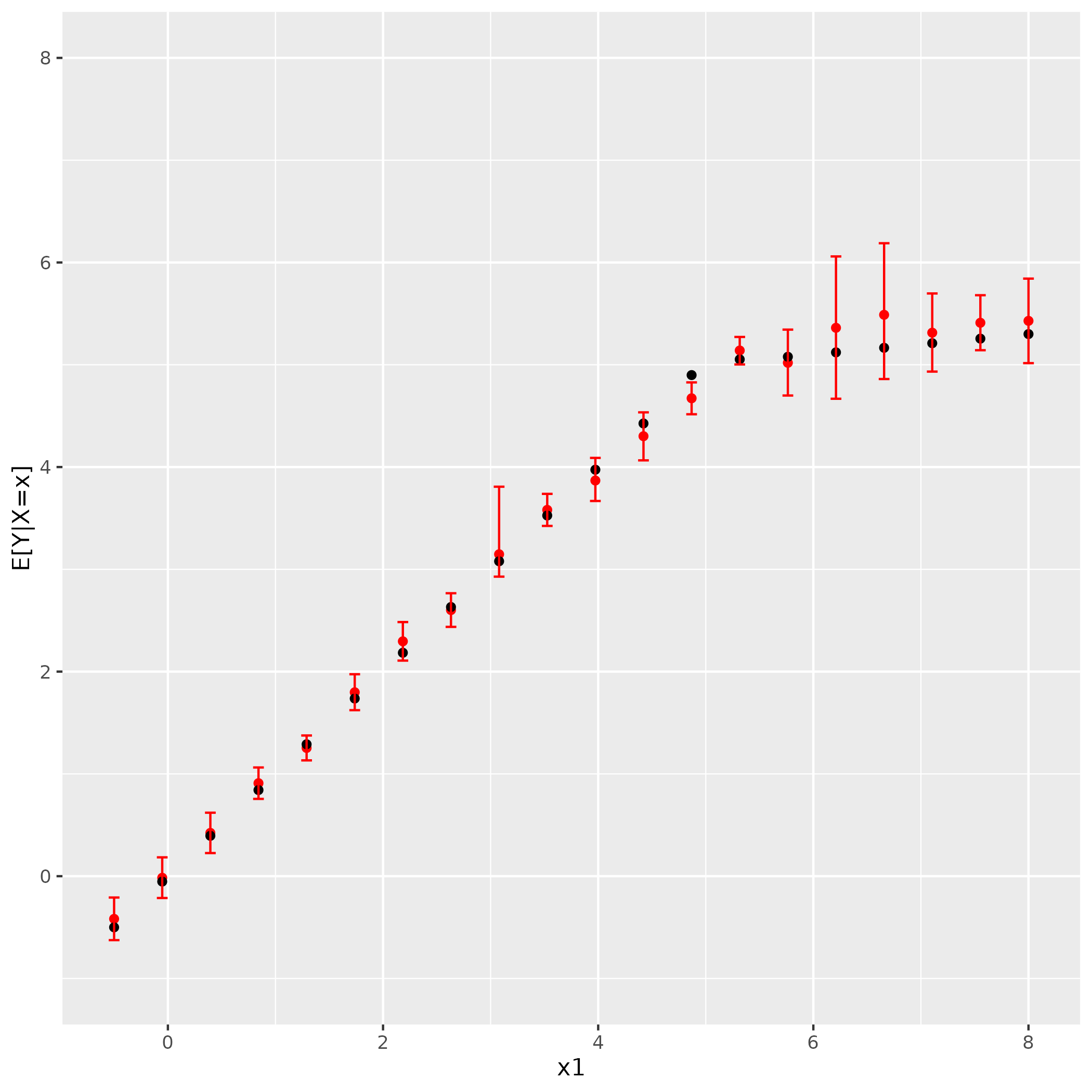} 
        \caption{BG, $p=15$, $N=10$, $M=50$} \label{fig:BG15}
    \end{subfigure}

    \vspace{0.5cm}
    \begin{subfigure}[h]{0.32\textwidth}
        \centering
        \includegraphics[width=\linewidth]{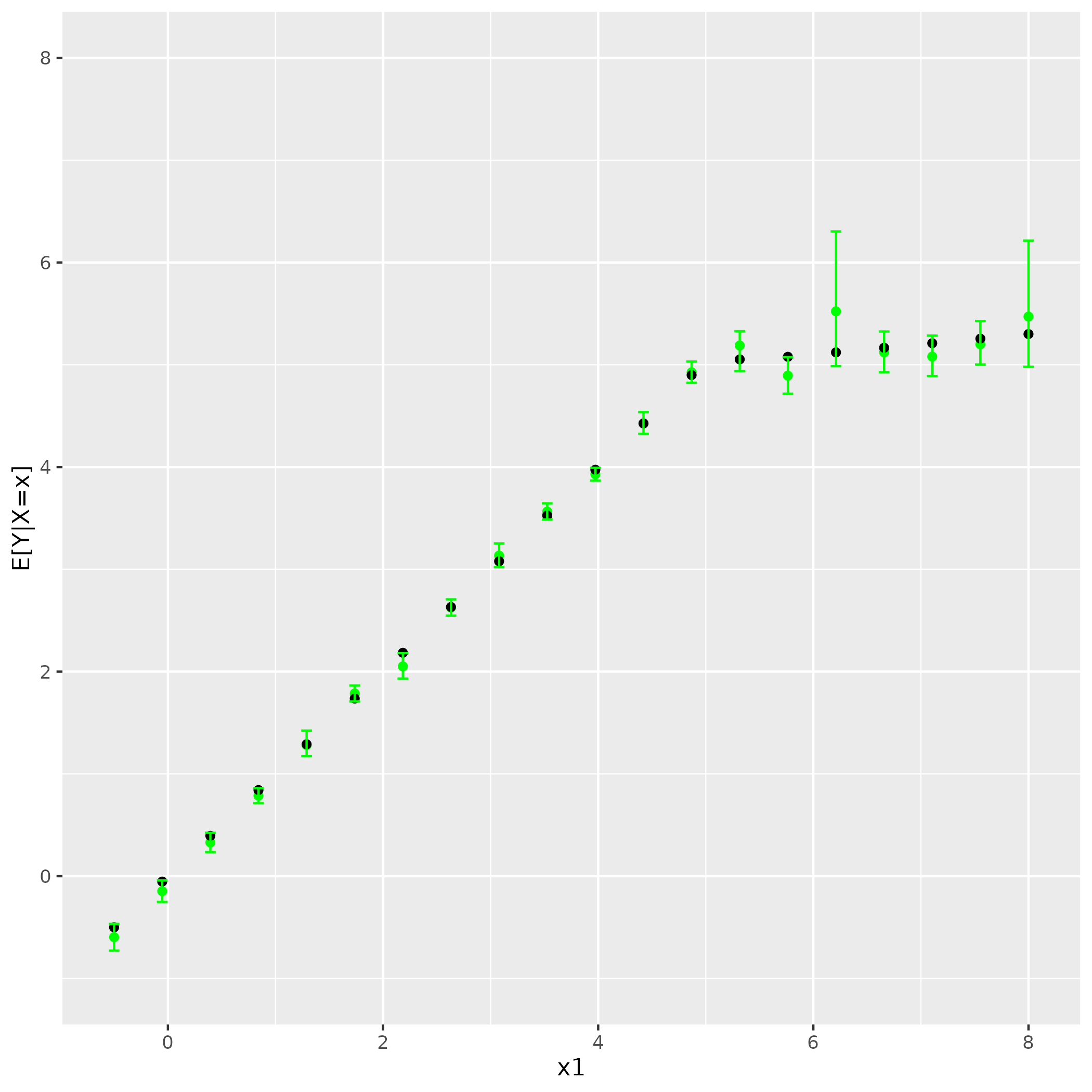} 
        \caption{BG, $p=5$, $N=7$, $M=33$} \label{fig:BGv5}
    \end{subfigure}
    \hfill
    \begin{subfigure}[h]{0.33\textwidth}
        \centering
        \includegraphics[width=\linewidth]{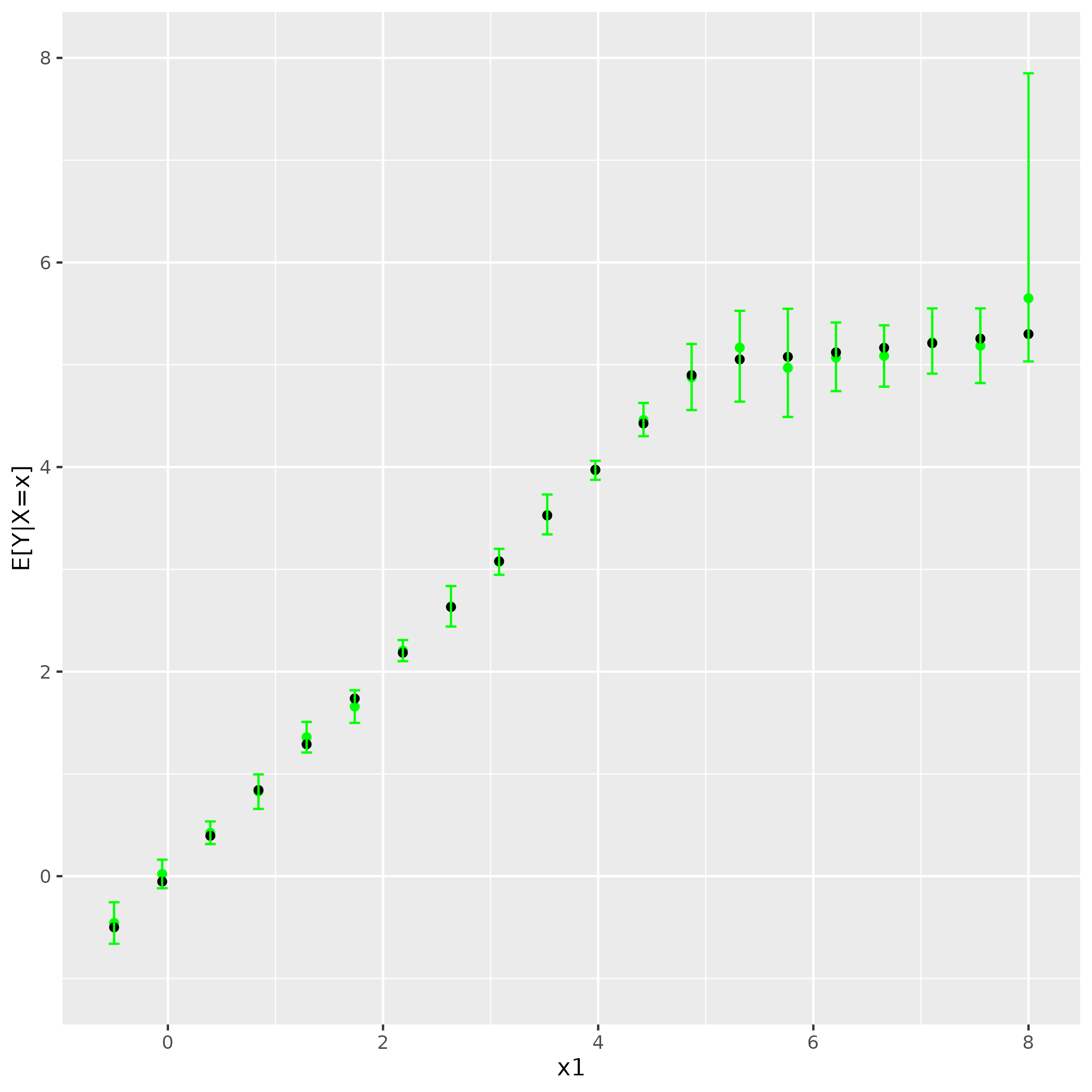} 
        \caption{BG, $p=10$, $N=9$, $M=23$} \label{fig:BGv10}
    \end{subfigure}
    \hfill
    \begin{subfigure}[h]{0.33\textwidth}
        \centering
        \includegraphics[width=\linewidth]{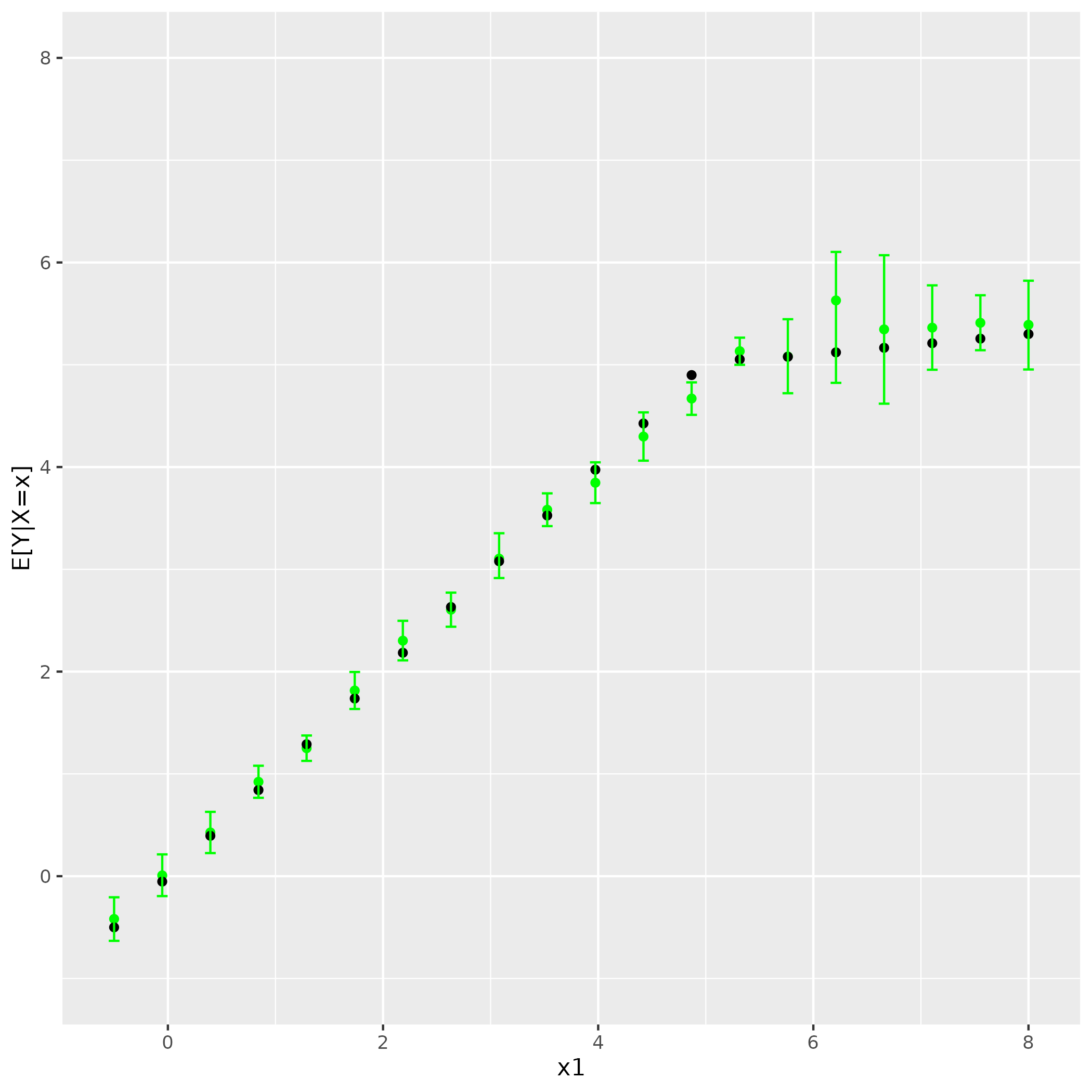} 
        \caption{BG, $p=15$, $N=7$, $M=21$} \label{fig:BGv15}
    \end{subfigure}
    \caption{$E[Y|X=x]$ for 20 new subjects plotted against each subject's value of $x_1$. Estimates of $E[Y|X=x]$ and 95\% credible intervals are shown in red for the blocked Gibbs sampler with fixed larger truncation values and green for the blocked Gibbs sampler with varying smaller truncation values. The true values of $E[Y|X=x]$ are shown in black.}
\end{figure}

Next, to assess whether the blocked Gibbs sampler with the minimum truncation values that provide adequately small error bounds is as accurate as the blocked Gibbs sampler with larger truncation values, we simulate 10 datasets and draw a test set of $m = 200$ new subjects for $p \in \{5,10,15\}$. We compute the true predictions for each new subject $E[Y_{n+i}|x_{n+i}]$ and compare them to the estimated predictions calculated using the estimates from each of the 10 datasets using the blocked Gibbs sampler with larger fixed truncation values and the blocked Gibbs sampler with small truncation values. Table \ref{tab:pred} shows the mean absolute ($\hat{l}_1$) and mean squared ($\hat{l}_2$) prediction error using each algorithm for each of the test sets. For each number of covariates and each algorithm, we calculate the prediction error estimates from each of the 10 datasets, and we obtain the values in Table \ref{tab:pred} by calculating the mean of the prediction error estimates over the 10 datasets. MC errors for these estimates can be found in Supplemental Table 1; the differences are well within the MC error. We can see that the truncation has essentially no effect on the prediction error.

\begin{table}
    \centering
    \begin{tabular}{ccccccc}
         \toprule
          & \multicolumn{2}{c}{$p=5$} & \multicolumn{2}{c}{$p=10$} & \multicolumn{2}{c}{$p=15$} \\
         \cmidrule{2-7}
          & $\hat{l}_1$ & $\hat{l}_2$ & $\hat{l}_1$ & $\hat{l}_2$ & $\hat{l}_1$ & $\hat{l}_2$ \\
         \midrule
         BG sampler, $N=10$, $M=50$ & 0.092 & 0.020 & 0.120 & 0.038 & 0.178 & 0.065 \\
         BG sampler, varying $N$ and $M$ & 0.089 & 0.021 & 0.118 & 0.031 & 0.162 & 0.061 \\
         \bottomrule
    \end{tabular}
    \caption{Estimated prediction error averaged over 10 simulations using each sampler for various numbers of covariates ($p$).}
    \label{tab:pred}
\end{table}

\subsection{Mixing}

We expect the blocked Gibbs sampler to have better mixing properties than the Polya urn Gibbs sampler since the blocked Gibbs sampler allows us to update blocks of parameters at a time, in contrast to the Polya urn Gibbs sampler requiring us to update the cluster assignment variables one subject at a time. In order to compare mixing between the blocked Gibbs sampler and the Polya urn Gibbs sampler, we use a similar method as Ishwaran and James \cite{ishwaran_gibbs_2001} used to compare Gibbs samplers for finite mixture models.

For each of $p=5$, 10, and 15, we again used the 10 simulated datasets of $n=200$ subjects. We used the same process described above to determine the truncation values $N$ and $M$ for each of the simulated samples necessary to obtain error bounds less than or equal to 0.01 using estimates of the concentration parameters from a short run of 10,000 iterations of the blocked Gibbs sampler. For each of the simulated samples, we examined 50,000-iteration runs of the blocked Gibbs sampler with a burn in period of 10,000 iterations with $N=10$ and $M=50$ and with smaller $N$ and $M$ and the Polya urn Gibbs sampler. Following the 10,000-iteration burn in period, for every 100 iterations we calculated the mean and the 0.025, 0.25, 0.75, and 0.975 quantiles of the estimates of $E[Y|X=x]$ for each of the 200 subjects. We then calculated the means and standard deviations of these summary statistics for each subject over the 400 batches of 100 iterations and averaged over the subjects to obtain averaged mean values and averaged standard deviations. Finally, we averaged these values over each of the ten simulated samples. These values are shown in Table \ref{tab:mix}. 

Both blocked Gibbs samplers, one with larger set $N$ and $M$ and one with smaller $N$ and $M$ depending on preliminary estimates of $\alpha^\theta$ and $\alpha^{\psi|\theta}_{max}$, generally have better mixing than the Polya urn Gibbs sampler as indicated by smaller standard deviations. For the 0.975 quantile for $p=5$ and $p=10$, the Polya urn Gibbs sampler performs similarly to the two blocked Gibbs samplers, but there are no cases in which the Polya urn Gibbs sampler performs substantially better than either of the blocked Gibbs samplers. It also appears that the improvement in mixing of the blocked Gibbs sampler with larger set $N$ and $M$ over the Polya urn Gibbs sampler may increase with increasing numbers of covariates since the largest difference between the average standard deviations for these two samplers occurs for $p=15$. For $p=15$, the blocked Gibbs sampler with larger set $N$ and $M$ also has better mixing than the blocked Gibbs sampler with smaller $N$ and $M$ in every case; however, the latter is clearly more computationally efficient per MCMC iteration.

\begin{table}
    \centering
    \begin{tabular}{c|c|cc|cc|cc}
         \toprule
          \multicolumn{2}{c}{}  & \multicolumn{2}{c|}{BG sampler} & \multicolumn{2}{c|}{BG sampler} & \multicolumn{2}{c}{PG sampler} \\
         \multicolumn{2}{c}{}  & \multicolumn{2}{c|}{$N=10$, $M=50$} & \multicolumn{2}{c|}{smaller $N$ and $M$} & \multicolumn{2}{c}{} \\
         \midrule
          $p$ & Statistic & Mean & SD & Mean & SD & Mean & SD \\
          \midrule
         \multirow{5}{*}{5} & 0.025 quantile & 3.60 & 0.024 & 3.60 & 0.024 & 3.60 & 0.028\\
& 0.25 quantile & 3.45 & 0.031 & 3.45 & 0.030 & 3.44 & 0.038\\
& mean & 3.54 & 0.025 & 3.54 & 0.025 & 3.55 & 0.032\\
& 0.75 quantile & 3.65 & 0.029 & 3.65 & 0.029 & 3.65 & 0.033\\
& 0.975 quantile & 3.76 & 0.048 & 3.76 & 0.049 & 3.75 & 0.046\\
\midrule
\multirow{5}{*}{10} & 0.025 quantile & 3.64 & 0.033 & 3.64 & 0.031 & 3.64 & 0.034\\
& 0.25 quantile & 3.44 & 0.044 & 3.45 & 0.044 & 3.45 & 0.046\\
& mean & 3.57 & 0.035 & 3.57 & 0.034 & 3.58 & 0.037\\
& 0.75 quantile & 3.71 & 0.036 & 3.71 & 0.035 & 3.71 & 0.038\\
& 0.975 quantile & 3.84 & 0.052 & 3.84 & 0.051 & 3.83 & 0.051\\
\midrule
\multirow{5}{*}{15} & 0.025 quantile & 3.73 & 0.026 & 3.73 & 0.035 & 3.73 & 0.038\\
& 0.25 quantile & 3.47 & 0.046 & 3.48 & 0.050 & 3.52 & 0.050\\
& mean & 3.64 & 0.030 & 3.64 & 0.038 & 3.65 & 0.040\\
& 0.75 quantile & 3.82 & 0.030 & 3.81 & 0.038 & 3.80 & 0.041\\
& 0.975 quantile & 3.99 & 0.047 & 3.97 & 0.051 & 3.94 & 0.053\\
\bottomrule
\end{tabular}
    \caption{Averages of ten simulated datasets of mean values and standard deviations over 400 batches, each of size 100 iterations, for the blocked Gibbs samplers and the Polya urn sampler.}
    \label{tab:mix}
\end{table}

\section{Discussion}
In this paper, we have introduced a finite mixture truncation approximation for the EDP prior and proposed a blocked Gibbs sampler for an EDPM of normals using the truncation approximation. We have shown that the blocked Gibbs sampler with $N$ and $M$ chosen by the derived error bounds  yields similar accuracy to the blocked Gibbs sampler with arbitrarily large $N$ and $M$ and that both of these can offer improved mixing over 
a Polya urn Gibbs sampler using an EDP prior with no truncation approximation. 
We are developing an R package to implement the blocked Gibbs sampler. However, we also note that the truncation approximation allows EDPMs to be fit in \textit{rstan} or \textit{rjags}. Further improvements in mixing and computational speed could be obtained by varying the truncation value $M_k$ within each $y$-cluster depending on the values of $\alpha_k^{\psi|\theta}$. This would allow the use of fewer $x$-clusters in the truncation approximation within $y$-clusters with smaller concentration parameters $\alpha_k^{\psi|\theta}$. Error bounds would need to be derived for this case.

\section*{Acknowledgments}

Natalie Burns and Michael J. Daniels were partially supported by NIH R01 HL166234.

\bibliographystyle{plain}
\bibliography{EDPPaperBib.bib}

\section*{Appendix A. Proofs}

\begin{proof}[Proof of Theorem 1] Following a similar approach to that of Ishwaran and James \cite{ishwaran_approximate_2002} in their proof of the $\mathcal{L}_1$ error bound for the truncation approximation for a Dirichlet process, we can first integrate over $P$ in order to write $m_{NM}$ and $m_\infty$ in terms of the distributions for $(\mathbf{y},\mathbf{x})$ under $\mathcal{P}_{NM}$ and $EDP(\alpha^\theta, \alpha^\psi, P_0)$, and call these two distributions $\pi_{NM}(\mathbf{y},\mathbf{x})$ and $\pi_{\infty}(\mathbf{y},\mathbf{x})$.  Then we have
\begin{align*}
    \int |m_{NM}(\mathbf{y},\mathbf{x}) - m_\infty(\mathbf{y},\mathbf{x})| &d(\mathbf{y},\mathbf{x}) \\
    &= \int\left|\int\prod_{i=1}^nf_y(y_i|x_i, \theta_i)f_x(x_i|\psi_i)(d\pi_{NM}(\boldsymbol{\theta},\boldsymbol{\psi}) - d\pi_{\infty}(\boldsymbol{\theta},\boldsymbol{\psi}))\right|d(\mathbf{y},\mathbf{x}) \\
    &\leq \int\int\prod_{i=1}^nf_y(y_i|x_i, \theta_i)f_x(x_i|\psi_i)\left|d\pi_{NM}(\boldsymbol{\theta},\boldsymbol{\psi}) - d\pi_{\infty}(\boldsymbol{\theta},\boldsymbol{\psi})\right|d(\mathbf{y},\mathbf{x}) \\
    &= \int\int\prod_{i=1}^nf_y(y_i|x_i, \theta_i)f_x(x_i|\psi_i)d(\mathbf{y},\mathbf{x})\left|d\pi_{NM}(\boldsymbol{\theta},\boldsymbol{\psi}) - d\pi_{\infty}(\boldsymbol{\theta},\boldsymbol{\psi})\right| \\
    &= \int\left|d\pi_{NM}(\boldsymbol{\theta},\boldsymbol{\psi}) - d\pi_{\infty}(\boldsymbol{\theta},\boldsymbol{\psi})\right| = 2D(\pi_{NM},\pi_{\infty}).
\end{align*}

Now let $K_i$ be the $\theta$-cluster from which the $i$th observation is drawn and $J_i$ be the $\psi$-cluster within that $\theta$-cluster from which the $i$th observation is drawn. We can write $\left(\theta_i,\psi_i\right) = \left(\theta_{K_i}^*,\psi_{J_i|K_i}^*\right)$, and when $K_i<N$ and $J_i<M$, $\left(\theta_i,\psi_i\right)$ under $\pi_{NM}$ is identical to $\left(\theta_i,\psi_i\right)$ under $\pi_{\infty}$.  Thus, we have
\begin{align*}
    D(\pi_{NM},\pi_{\infty}) &= \sup_A\left|\pi_{NM}(A) - \pi_{\infty}(A)\right| \\
    &= \sup_A\left|\pi_{NM}(A\cap \{K_i<N,J_i<M,i=1,\ldots,n\}^C) \right. \\
    &\phantom{\leq\sup_A\Bigg[} \left. - \pi_{\infty}(A\cap \{K_i<N,J_i<M,i=1,\ldots,n\}^C)\right| \\
    &\leq \sup_A\left[\pi_{NM}(A\cap \{K_i<N,J_i<M,i=1,\ldots,n\}^C)\right] \\
    &\phantom{\leq\sup_A} + \sup_A\left[\pi_{\infty}(A\cap \{K_i<N,J_i<M,i=1,\ldots,n\}^C)\right] \\
    &= 2\left(1 - \pi_{NM}\{K_i<N,J_i<M,i=1,\ldots,n\}\right) \\
    &= 2\left[1 - E\left\{\left(\sum_{k=1}^{N-1}p_k^{\theta}\sum_{j=1}^{M-1}p_{j|k}^{\psi}\right)^n\right\}\right].
\end{align*}

Then we have
\begin{align}
    E\left\{\left(\sum_{k=1}^{N-1}p_k^{\theta}\sum_{j=1}^{M-1}p_{j|k}^{\psi}\right)^n\right\} &= E\left\{\left(1 - \left(p_N^\theta + \sum_{k=1}^{N-1}p_k^\theta p_{M|k}^{\psi}\right)\right)^n\right\} \nonumber \\
    &\approx E\left\{1 - n\left(p_N^\theta + \sum_{k=1}^{N-1}p_k^\theta p_{M|k}^{\psi}\right)\right\}
\end{align}
since $\left|- \left(p_N^\theta + \sum_{k=1}^{N-1}p_k^\theta p_{M|k}^{\psi}\right)\right| < 1$ with probability 1 and $\Pr\left(\left|-n\left(p_N^\theta + \sum_{k=1}^{N-1}p_k^\theta p_{M|k}^{\psi}\right)\right| \ll 1\right) \approx 1$ for sufficiently large $N$ and $M$.  To show the final approximation in the previous sentence, let $\epsilon \ll 1$.  Then, we have
\begin{align*}
    \Pr\left(\left|-n\left(p_N^\theta + \sum_{k=1}^{N-1}p_k^\theta p_{M|k}^{\psi}\right)\right| \ll 1\right) &= \Pr\left(n\left(p_N^\theta + \sum_{k=1}^{N-1}p_k^\theta p_{M|k}^{\psi}\right) \ll 1\right) \\
    &> \Pr\left(n\left(p_N^\theta + \sum_{k=1}^{N-1}p_k^\theta p_{M|k}^{\psi}\right) < \epsilon\right) \\
    &= 1 - \Pr\left(p_N^\theta + \sum_{k=1}^{N-1}p_k^\theta p_{M|k}^{\psi} \geq \epsilon/n\right) \\
    &\geq 1 - \frac{E\left(p_N^\theta + \sum_{k=1}^{N-1}p_k^\theta p_{M|k}^{\psi}\right)}{\epsilon / n} \\ 
    &= 1 - \frac{n}{\epsilon}\left[E\left(p_N^\theta\right) + \sum_{k=1}^{N-1}E\left(p_k^\theta p_{M|k}^\psi\right) \right] \\
    &> 1 - \frac{n}{\epsilon}\left[E\left(p_N^\theta\right) + \sum_{k=1}^{N-1}E\left( p_{M|k}^\psi\right) \right],
\end{align*}
with the last inequality holding because $p_k^\theta < 1$ with probability one for all $k=1,\ldots,N-1$.  Finally,
\begin{align*}
    1 - \frac{n}{\epsilon}\left[E\left(p_N^\theta\right) + \sum_{k=1}^{N-1}E\left( p_{M|k}^\psi\right) \right] &= 1 - \frac{n}{\epsilon}\left[E\left(\exp{\left\{-\frac{1}{\alpha^\theta}\sum_{k=1}^{N-1}E_k^\theta\right\}}\right) \right. \\
    &\phantom{= 1 - \frac{n}{\epsilon}[E(\exp} \left. + \sum_{k=1}^{N-1}E\left(\exp{\left\{-\frac{1}{\alpha^\psi}\sum_{j=1}^{M-1}E_{j|k}^\psi\right\}}\right) \right]
\end{align*}
where $E_k^\theta \stackrel{\text{iid}}{\sim} exp(1)$ and $E_{j|k}^\psi \stackrel{\text{iid}}{\sim} exp(1)$.  Then, $\sum_{k=1}^{N-1}E_k^\theta \sim Gamma(N-1,1)$ and $\sum_{j=1}^{M-1}E_{j|k}^\psi \sim Gamma(M-1,1)$, so
\begin{align*}
    \Pr\left(\left|-n\left(p_N^\theta + \sum_{k=1}^{N-1}p_k^\theta p_{M|k}^{\psi}\right)\right| \ll 1\right) &> 1 - \frac{n}{\epsilon}\left[\left(1+\frac{1}{\alpha^\theta}\right)^{-(N-1)} + \sum_{k=1}^{N-1}\left(1+\frac{1}{\alpha^\psi}\right)^{-(M-1)}\right]
\end{align*}
Thus, we can make $\Pr\left(\left|-n\left(p_N^\theta + \sum_{k=1}^{N-1}p_k^\theta p_{M|k}^{\psi}\right)\right| \ll 1\right)$ arbitrarily close to 1 by making $N$ and $M$ sufficiently large.

Continuing from (4), we have
\begin{align*}
    E(p_N^\theta) &= E\left[\left(1-V_1^\theta\right)\left(1-V_2^\theta\right)\cdots\left(1-V_{N-1}^\theta\right)\right] \\
    &= E\left[\exp{\left\{-\frac{1}{\alpha^\theta}\sum_{k=1}^{N-1}E_k^\theta\right\}}\right],
\end{align*}
where again $E_k^\theta \stackrel{\text{iid}}{\sim} exp(1)$, and thus,
\begin{align}
    E(p_N^\theta)
    &\approx \exp{\left\{-\frac{N-1}{\alpha^\theta}\right\}}.
\end{align}
We also have
\begin{align*}
    E\left(\sum_{k=1}^{N-1}p_k^\theta p_{M|k}^{\psi}\right) &= EE\left(\sum_{k=1}^{N-1}p_k^\theta p_{M|k}^{\psi} \Bigg| p_1^\theta,\ldots,p_{N-1}^\theta,\theta_1^*,\ldots,\theta_{N-1}^*\right) \\
    &= E\left[\sum_{k=1}^{N-1}p_k^\theta E\left(p_{M|k}^{\psi} \Bigg| p_1^\theta,\ldots,p_{N-1}^\theta,\theta_1^*,\ldots,\theta_{N-1}^*\right)\right].
\end{align*}
Then using the same process as for $E(p_N^\theta)$ and noting that the probabilities of the $\psi$-clusters do not depend on the $\theta$-cluster parameters or the $\theta$-cluster probabilities, we have
\begin{align*}
    E\left(p_{M|k}^{\psi} \Bigg| p_1^\theta,\ldots,p_{N-1}^\theta,\theta_1^*,\ldots,\theta_{N-1}^*\right) &= E\left(p_{M|k}^{\psi}\right) \\
    &= E\left[\left(1-V_{1|k}^\psi\right)\left(1-V_{2|k}^\psi\right)\cdots\left(1-V_{M-1|k}^\psi\right)\right] \\
    &= E\left[\exp{\left\{-\frac{1}{\alpha^\psi}\sum_{j=1}^{M-1}E_{j|k}^\psi\right\}}\right],
\end{align*}
where again $E_{j|k}^\psi \stackrel{\text{iid}}{\sim} exp(1)$, and thus,
\begin{align}
    E(p_{M|k}^\psi) &\approx \exp{\left\{-\frac{M-1}{\alpha^\psi}\right\}},
\end{align}
so
\begin{align}
    E\left(\sum_{k=1}^{N-1}p_k^\theta p_{M|k}^{\psi}\right) &\approx E\left[\sum_{k=1}^{N-1}p_k^\theta \exp{\left\{-\frac{M-1}{\alpha^\psi}\right\}}\right] \nonumber \\
    &= \exp{\left\{-\frac{M-1}{\alpha^\psi}\right\}}E\left(\sum_{k=1}^{N-1}p_k^\theta\right) \nonumber \\
    &= \exp{\left\{-\frac{M-1}{\alpha^\psi}\right\}}E\left(1-p_N^\theta\right) \nonumber \\
    &\approx \exp{\left\{-\frac{M-1}{\alpha^\psi}\right\}}\left(1 - \exp{\left\{-\frac{N-1}{\alpha^\theta}\right\}}\right).
\end{align}

Finally, plugging the approximations from (5) and (7) into (4), this gives us 
\begin{align}
    E\left\{\left(\sum_{k=1}^{N-1}p_k^{\theta}\sum_{j=1}^{M-1}p_{j|k}^{\psi}\right)^n\right\} &\approx 1 - n\left(\exp{\left\{-\frac{N-1}{\alpha^\theta}\right\}} + \exp{\left\{-\frac{M-1}{\alpha^\psi}\right\}}\left(1 - \exp{\left\{-\frac{N-1}{\alpha^\theta}\right\}}\right)\right).
\end{align}

\end{proof}
\pagebreak

\begin{proof}[Details of Corollary 1]
Aside from replacing $\alpha^\psi$ with $\alpha^{\psi|\theta}_k$, the proof of Corollary 1 is identical to the proof of Theorem 1 up until (6), which for the case in which $\alpha^\psi$ depends on with which $\theta$-cluster it is associated, we have
\begin{align*}
    E(p_{M|k}^\psi) &\approx \exp{\left\{-\frac{M-1}{\alpha^{\psi|\theta}_k}\right\}},
\end{align*}
so
\begin{align}
    E\left(\sum_{k=1}^{N-1}p_k^\theta p_{M|k}^{\psi}\right) &\approx E\left[\sum_{k=1}^{N-1}p_k^\theta \exp{\left\{-\frac{M-1}{\alpha^{\psi|\theta}_k}\right\}}\right] \nonumber \\
    &\leq \exp{\left\{-\frac{M-1}{\alpha^{\psi|\theta}_\text{max}}\right\}}E\left(\sum_{k=1}^{N-1}p_k^\theta\right) \nonumber \\
    &= \exp{\left\{-\frac{M-1}{\alpha^{\psi|\theta}_\text{max}}\right\}}E\left(1-p_N^\theta\right) \nonumber \\
    &\approx \exp{\left\{-\frac{M-1}{\alpha^{\psi|\theta}_\text{max}}\right\}}\left(1 - \exp{\left\{-\frac{N-1}{\alpha^\theta}\right\}}\right),
\end{align}
where $\alpha^{\psi|\theta}_\text{max}$ is the largest of $\alpha^{\psi|\theta}_1,\ldots,\alpha^{\psi|\theta}_N$.
\end{proof}

\begin{proof}[Proof of Corollary 2]
Again, following a similar approach to that of Ishwaran and James \cite{ishwaran_approximate_2002} in their proof of the error bound for the posterior of the classification variables under the truncation approximation for a Dirichlet process, let $\mathcal{K}_r=\{1,2,\ldots,r\}^n$ and $\mathcal{J}_m=\{1,2,\ldots,m\}^n$ for $r=1,2,\ldots$ and $m=1,2,\ldots$.  We have
\begin{align}
    \sum_{\mathbf{K}\in\mathcal{K}_\infty}\sum_{\mathbf{J}\in\mathcal{J}_\infty}\left|\pi_{NM}(\mathbf{K},\mathbf{J} \right. &| \left. \mathbf{y},\mathbf{x}) - \pi_{\infty}(\mathbf{K},\mathbf{J}|\mathbf{y},\mathbf{x})\right| \nonumber \\
    &= \sum_{\mathbf{K}\in\mathcal{K}_\infty}\left[\sum_{\mathbf{J}\in\mathcal{J}_M}\left|\pi_{NM}(\mathbf{K},\mathbf{J}|\mathbf{y},\mathbf{x}) - \pi_{\infty}(\mathbf{K},\mathbf{J}|\mathbf{y},\mathbf{x})\right|\right. \nonumber \\
    &\phantom{\sum_{\mathbf{K}\in\mathcal{K}_\infty}\Bigg[\sum_{\mathbf{J}\in\mathcal{J}_M}} \left. + \sum_{\mathbf{J}\in\mathcal{J}_\infty - \mathcal{J}_M}\left|\pi_{NM}(\mathbf{K},\mathbf{J}|\mathbf{y},\mathbf{x}) - \pi_{\infty}(\mathbf{K},\mathbf{J}|\mathbf{y},\mathbf{x})\right|\right] \nonumber \\
    &= \sum_{\mathbf{K}\in\mathcal{K}_N}\left[\sum_{\mathbf{J}\in\mathcal{J}_M}\left|\pi_{NM}(\mathbf{K},\mathbf{J}|\mathbf{y},\mathbf{x}) - \pi_{\infty}(\mathbf{K},\mathbf{J}|\mathbf{y},\mathbf{x})\right|\right. \nonumber \\
    &\phantom{\sum_{\mathbf{K}\in\mathcal{K}_\infty}\Bigg[\sum_{\mathbf{J}\in\mathcal{J}_M}} \left. + \sum_{\mathbf{J}\in\mathcal{J}_\infty - \mathcal{J}_M} \pi_{\infty}(\mathbf{K},\mathbf{J}|\mathbf{y},\mathbf{x})\right] \nonumber \\
    &\quad + \sum_{\mathbf{K}\in\mathcal{K}_\infty-\mathcal{K}_N}\left[\sum_{\mathbf{J}\in\mathcal{J}_M}\left|\pi_{NM}(\mathbf{K},\mathbf{J}|\mathbf{y},\mathbf{x}) - \pi_{\infty}(\mathbf{K},\mathbf{J}|\mathbf{y},\mathbf{x})\right|\right. \nonumber \\
    &\phantom{\sum_{\mathbf{K}\in\mathcal{K}_\infty}\Bigg[\sum_{\mathbf{J}\in\mathcal{J}_M}} \left. + \sum_{\mathbf{J}\in\mathcal{J}_\infty - \mathcal{J}_M} \pi_{\infty}(\mathbf{K},\mathbf{J}|\mathbf{y},\mathbf{x})\right] \nonumber \\
    &= \sum_{\mathbf{K}\in\mathcal{K}_N}\sum_{\mathbf{J}\in\mathcal{J}_M}\left|\pi_{NM}(\mathbf{K},\mathbf{J}|\mathbf{y},\mathbf{x}) - \pi_{\infty}(\mathbf{K},\mathbf{J}|\mathbf{y},\mathbf{x})\right| \nonumber \\
    &\phantom{\sum_{\mathbf{K}\in\mathcal{K}_N}} + \sum_{\mathbf{K}\in\mathcal{K}_N}\sum_{\mathbf{J}\in\mathcal{J}_\infty - \mathcal{J}_M}\pi_{\infty}(\mathbf{K},\mathbf{J}|\mathbf{y},\mathbf{x}) + \sum_{\mathbf{K}\in\mathcal{K}_\infty-\mathcal{K}_N}\sum_{\mathbf{J}\in\mathcal{J}_\infty}\pi_{\infty}(\mathbf{K},\mathbf{J}|\mathbf{y},\mathbf{x}).
\end{align}

We can write
\begin{align*}
    \pi_{NM}(\mathbf{K},\mathbf{J}|\mathbf{y},\mathbf{x}) &= \frac{\Pr_{NM}(\mathbf{K},\mathbf{J})m_{NM}(\mathbf{y},\mathbf{x}|\mathbf{K},\mathbf{J})}{m_{NM}(\mathbf{y},\mathbf{x})},\quad \mathbf{K}\in\mathcal{K}_N\,\, \text{and}\,\, \mathbf{J}\in\mathcal{J}_M
\end{align*}
where $\Pr_{NM}(\mathbf{K},\mathbf{J})$ is the prior for $(\mathbf{K},\mathbf{J})$ under $\mathcal{P}_{NM}$ and
\begin{align*}
    m_{NM}(\mathbf{y},\mathbf{x}|\mathbf{K},\mathbf{J}) &= \prod_{k\in\mathbf{K}^*,j\in\mathbf{J}^*}\left[\int_{\mathbb{R}\times\mathbb{R}^+\times\mathbb{R}\times\mathbb{R}^+}\left(\prod_{\{i:K_i=k,J_i=j\}}f_y(y_i|x_i, \theta)f_x(x_i|\psi)\right)dP_0(\theta,\psi)\right],
\end{align*}
where $\mathbf{K}^*$ are the unique values of $K_i$ and $\mathbf{J}^*$ are the unique values of $J_i$. Note that $m_{NM}(\mathbf{y},\mathbf{x}|\mathbf{K},\mathbf{J}) = m_{\infty}(\mathbf{y},\mathbf{x}|\mathbf{K},\mathbf{J})$ for $\mathbf{K}\in\mathcal{K}_N$ and $\mathbf{J}\in\mathcal{J}_M$ since all cluster parameters are the same for each observation that comes from one of the first $N$ $\theta$-clusters and one of the first $M$ $\psi$-clusters within each of the first $N$ $\theta$-clusters.  Also, $\Pr_{NM}(\mathbf{K},\mathbf{J}) = \Pr_{\infty}(\mathbf{K},\mathbf{J})$ for $\mathbf{K}\in\mathcal{K}_{N-1}$ and $\mathbf{J}\in\mathcal{J}_{M-1}$ since the probabilities of being in each of the first $N-1$ $\theta$-clusters and each of the first $M-1$ $\psi$-clusters within each of the first $N-1$ $\theta$-clusters are the same under $\mathcal{P}_{NM}$ and $EDP(\alpha^\theta, \alpha^\psi, P_0)$.  However, we do not necessarily have $\Pr_{NM}(\mathbf{K},\mathbf{J}) = \Pr_{\infty}(\mathbf{K},\mathbf{J})$ if $\mathbf{K}\in\mathcal{K}_{N}$ or $\mathbf{J}\in\mathcal{J}_{M}$.  Now, looking at the first sum in the right-hand side of (10), we have
\begin{align*}
    \sum_{\mathbf{K}\in\mathcal{K}_N} \sum_{\mathbf{J}\in\mathcal{J}_M}&\left|\pi_{NM}(\mathbf{K},\mathbf{J}|\mathbf{y},\mathbf{x}) - \pi_{\infty}(\mathbf{K},\mathbf{J}|\mathbf{y},\mathbf{x})\right|  \\
    &= \sum_{\mathbf{K}\in\mathcal{K}_N}\sum_{\mathbf{J}\in\mathcal{J}_M}\left|\frac{\Pr_{NM}(\mathbf{K},\mathbf{J})m_{NM}(\mathbf{y},\mathbf{x}|\mathbf{K},\mathbf{J})}{m_{NM}(\mathbf{y},\mathbf{x})} - \frac{\Pr_{\infty}(\mathbf{K},\mathbf{J})m_{\infty}(\mathbf{y},\mathbf{x}|\mathbf{K},\mathbf{J})}{m_{\infty}(\mathbf{y},\mathbf{x})}\right| \\
    &= \sum_{\mathbf{K}\in\mathcal{K}_N}\sum_{\mathbf{J}\in\mathcal{J}_M}\left|\frac{\Pr_{NM}(\mathbf{K},\mathbf{J})m_{NM}(\mathbf{y},\mathbf{x}|\mathbf{K},\mathbf{J})}{m_{NM}(\mathbf{y},\mathbf{x})} \right. \\
    &\phantom{\sum_{\mathbf{K}\in\mathcal{K}_N}} \left. - \frac{\Pr_{NM}(\mathbf{K},\mathbf{J})m_{NM}(\mathbf{y},\mathbf{x}|\mathbf{K},\mathbf{J}) + \left[\Pr_{\infty}(\mathbf{K},\mathbf{J})-\Pr_{NM}(\mathbf{K},\mathbf{J})\right]m_{NM}(\mathbf{y},\mathbf{x}|\mathbf{K},\mathbf{J})}{m_{\infty}(\mathbf{y},\mathbf{x})}\right| \\
    &\leq \left|1 - \frac{m_{NM}(\mathbf{y},\mathbf{x})}{m_{\infty}(\mathbf{y},\mathbf{x})}\right| \sum_{\mathbf{K}\in\mathcal{K}_N}\sum_{\mathbf{J}\in\mathcal{J}_M} \frac{\Pr_{NM}(\mathbf{K},\mathbf{J})m_{NM}(\mathbf{y},\mathbf{x}|\mathbf{K},\mathbf{J})}{m_{NM}(\mathbf{y},\mathbf{x})} \\
    &\phantom{\Bigg|1-} + \sum_{\mathbf{K}\in\mathcal{K}_N}\sum_{\mathbf{J}\in\mathcal{J}_M} \frac{ \left|\Pr_{\infty}(\mathbf{K},\mathbf{J})-\Pr_{NM}(\mathbf{K},\mathbf{J})\right|m_{NM}(\mathbf{y},\mathbf{x}|\mathbf{K},\mathbf{J})}{ m_{\infty}(\mathbf{y},\mathbf{x})} \\
    &= \left|1 - \frac{m_{NM}(\mathbf{y},\mathbf{x})}{m_{\infty}(\mathbf{y},\mathbf{x})}\right| \sum_{\mathbf{K}\in\mathcal{K}_N}\sum_{\mathbf{J}\in\mathcal{J}_M} \pi_{NM}(\mathbf{K},\mathbf{J}|\mathbf{y},\mathbf{x}) \\
    &\phantom{\Bigg|1-} + \sum_{\mathbf{K}\in\mathcal{K}_N-\mathcal{K}_{N-1}}\sum_{\mathbf{J}\in\mathcal{J}_M} \frac{ m_{NM}(\mathbf{y},\mathbf{x}|\mathbf{K},\mathbf{J})}{ m_{\infty}(\mathbf{y},\mathbf{x})}\left|\Pr{}_{\infty}(\mathbf{K},\mathbf{J})-\Pr{}_{NM}(\mathbf{K},\mathbf{J})\right| \\
    &\phantom{\Bigg|1-} + \sum_{\mathbf{K}\in\mathcal{K}_{N-1}}\sum_{\mathbf{J}\in\mathcal{J}_M-\mathcal{J}_{M-1}} \frac{ m_{NM}(\mathbf{y},\mathbf{x}|\mathbf{K},\mathbf{J})}{ m_{\infty}(\mathbf{y},\mathbf{x})}\left|\Pr{}_{\infty}(\mathbf{K},\mathbf{J})-\Pr{}_{NM}(\mathbf{K},\mathbf{J})\right| \\
    &\leq \left|1 - \frac{m_{NM}(\mathbf{y},\mathbf{x})}{m_{\infty}(\mathbf{y},\mathbf{x})}\right| + \sum_{\mathbf{K}\in\mathcal{K}_N-\mathcal{K}_{N-1}}\sum_{\mathbf{J}\in\mathcal{J}_M} \frac{ m_{NM}(\mathbf{y},\mathbf{x}|\mathbf{K},\mathbf{J})}{ m_{\infty}(\mathbf{y},\mathbf{x})}\left|\Pr{}_{\infty}(\mathbf{K},\mathbf{J})-\Pr{}_{NM}(\mathbf{K},\mathbf{J})\right| \\
    &\phantom{\Bigg|1-} + \sum_{\mathbf{K}\in\mathcal{K}_{N-1}}\sum_{\mathbf{J}\in\mathcal{J}_M-\mathcal{J}_{M-1}} \frac{ m_{NM}(\mathbf{y},\mathbf{x}|\mathbf{K},\mathbf{J})}{ m_{\infty}(\mathbf{y},\mathbf{x})}\left|\Pr{}_{\infty}(\mathbf{K},\mathbf{J})-\Pr{}_{NM}(\mathbf{K},\mathbf{J})\right|.
\end{align*}
Then, integrating with respect to $m_\infty(\mathbf{y},\mathbf{x})$, we have
\allowdisplaybreaks[0]
\begin{align}
    \int_{\mathbb{R}^n\times\mathbb{R}^n}\sum_{\mathbf{K}\in\mathcal{K}_N}\sum_{\mathbf{J}\in\mathcal{J}_M}&\left|\pi_{NM}(\mathbf{K},\mathbf{J}|\mathbf{y},\mathbf{x}) - \pi_{\infty}(\mathbf{K},\mathbf{J}|\mathbf{y},\mathbf{x})\right|d(\mathbf{y},\mathbf{x}) \nonumber  \\
    &\leq \int\left|m_{\infty}(\mathbf{y},\mathbf{x}) - m_{NM}(\mathbf{y},\mathbf{x})\right|d(\mathbf{y},\mathbf{x}) \nonumber \\
    &\phantom{\leq \int} + \int\sum_{\mathbf{K}\in\mathcal{K}_N-\mathcal{K}_{N-1}}\sum_{\mathbf{J}\in\mathcal{J}_M} m_{NM}(\mathbf{y},\mathbf{x}|\mathbf{K},\mathbf{J}) \left|\Pr{}_{\infty}(\mathbf{K},\mathbf{J})-\Pr{}_{NM}(\mathbf{K},\mathbf{J})\right| d(\mathbf{y},\mathbf{x}) \nonumber \\
    &\phantom{\leq \int} + \int\sum_{\mathbf{K}\in\mathcal{K}_{N-1}}\sum_{\mathbf{J}\in\mathcal{J}_M-\mathcal{J}_{M-1}} m_{NM}(\mathbf{y},\mathbf{x}|\mathbf{K},\mathbf{J}) \left|\Pr{}_{\infty}(\mathbf{K},\mathbf{J})-\Pr{}_{NM}(\mathbf{K},\mathbf{J})\right| d(\mathbf{y},\mathbf{x}) \nonumber \\
    &= \int\left|m_{\infty}(\mathbf{y},\mathbf{x}) - m_{NM}(\mathbf{y},\mathbf{x})\right|d(\mathbf{y},\mathbf{x}) \nonumber \\
    &\phantom{= \int} + \sum_{\mathbf{K}\in\mathcal{K}_N-\mathcal{K}_{N-1}}\sum_{\mathbf{J}\in\mathcal{J}_M}  \left|\Pr{}_{\infty}(\mathbf{K},\mathbf{J})-\Pr{}_{NM}(\mathbf{K},\mathbf{J})\right| \nonumber \\
    &\phantom{= \int} + \sum_{\mathbf{K}\in\mathcal{K}_{N-1}}\sum_{\mathbf{J}\in\mathcal{J}_M-\mathcal{J}_{M-1}} \left|\Pr{}_{\infty}(\mathbf{K},\mathbf{J})-\Pr{}_{NM}(\mathbf{K},\mathbf{J})\right|.
\end{align}
\allowdisplaybreaks

Now, $\int\left|m_{\infty}(\mathbf{y},\mathbf{x}) - m_{NM}(\mathbf{y},\mathbf{x})\right|d(\mathbf{y},\mathbf{x}) = O\left(n\left(\exp{\left\{-\frac{N-1}{\alpha^\theta}\right\}} + \exp{\left\{-\frac{M-1}{\alpha^\psi}\right\}}\left(1 - \exp{\left\{-\frac{N-1}{\alpha^\theta}\right\}}\right)\right)\right)$ by Theorem 1.  We also have both $\sum_{\mathbf{K}\in\mathcal{K}_N-\mathcal{K}_{N-1}}\sum_{\mathbf{J}\in\mathcal{J}_M}  \left|\Pr{}_{\infty}(\mathbf{K},\mathbf{J})-\Pr{}_{NM}(\mathbf{K},\mathbf{J})\right|$ and

\noindent $\sum_{\mathbf{K}\in\mathcal{K}_{N-1}}\sum_{\mathbf{J}\in\mathcal{J}_M-\mathcal{J}_{M-1}} \left|\Pr{}_{\infty}(\mathbf{K},\mathbf{J})-\Pr{}_{NM}(\mathbf{K},\mathbf{J})\right|$ bounded by 

\noindent $\sum_{\mathbf{K}\in\mathcal{K}_{\infty}}\sum_{\mathbf{J}\in\mathcal{J}_\infty} \left|\Pr{}_{\infty}(\mathbf{K},\mathbf{J})-\Pr{}_{NM}(\mathbf{K},\mathbf{J})\right|$, and 
\begin{align*}
    \sum_{\mathbf{K}\in\mathcal{K}_{\infty}}\sum_{\mathbf{J}\in\mathcal{J}_\infty} \left|\Pr{}_{\infty}(\mathbf{K},\mathbf{J})-\Pr{}_{NM}(\mathbf{K},\mathbf{J})\right| &= 2D(\Pr{}_{\infty},\Pr{}_{NM}) \\
    &\leq 4\left(1 - E\left[\left(\sum_{k=1}^{N-1}p_k^\theta\sum_{j=1}^{M-1}p_{j|k}^\psi\right)^n\right]\right) \\
    &\approx 4n\left[\exp{\left\{-\frac{N-1}{\alpha^\theta}\right\}} \right. \\
    &\phantom{\approx 4n[\exp} \left. + \exp{\left\{-\frac{M-1}{\alpha^\psi}\right\}}\left(1 - \exp{\left\{-\frac{N-1}{\alpha^\theta}\right\}}\right)\right],
\end{align*}
by a similar argument as that used in the proof of Theorem 1, so both the second and third terms in (11) are also $O\left(n\left(\exp{\left\{-\frac{N-1}{\alpha^\theta}\right\}} + \exp{\left\{-\frac{M-1}{\alpha^\psi}\right\}}\left(1 - \exp{\left\{-\frac{N-1}{\alpha^\theta}\right\}}\right)\right)\right)$.

Finally, we integrate the second and third sums on the right-hand side of (10) with respect to $m_\infty(\mathbf{y},\mathbf{x})$ to obtain $\sum_{\mathbf{K}\in\mathcal{K}_N}\sum_{\mathbf{J}\in\mathcal{J}_\infty - \mathcal{J}_M}\Pr_{\infty}(\mathbf{K},\mathbf{J})$ and $\sum_{\mathbf{K}\in\mathcal{K}_\infty-\mathcal{K}_N}\sum_{\mathbf{J}\in\mathcal{J}_\infty}\Pr_{\infty}(\mathbf{K},\mathbf{J})$, respectively.  We have 
\begin{align*}
    \sum_{\mathbf{K}\in\mathcal{K}_N}\sum_{\mathbf{J}\in\mathcal{J}_\infty - \mathcal{J}_M}\Pr{}_{\infty}(\mathbf{K},\mathbf{J}) &+ \sum_{\mathbf{K}\in\mathcal{K}_\infty-\mathcal{K}_N}\sum_{\mathbf{J}\in\mathcal{J}_\infty}\Pr{}_{\infty}(\mathbf{K},\mathbf{J}) \\
    &= \Pr{}_\infty\left\{K_i\leq N, J_i>M, i=1,\ldots,n\right\} + \Pr{}_\infty\left\{K_i> N, i=1,\ldots,n\right\} \\
    &= 1 - \Pr{}_\infty\left\{K_i\leq N, J_i\leq M, i=1,\ldots,n\right\} \\
    &\leq 1 - \Pr{}_\infty\left\{K_i< N, J_i< M, i=1,\ldots,n\right\} \\
    &= 1 - \Pr{}_{NM}\left\{K_i< N, J_i< M, i=1,\ldots,n\right\} \\
    &\approx n\left[\exp{\left\{-\frac{N-1}{\alpha^\theta}\right\}} + \exp{\left\{-\frac{M-1}{\alpha^\psi}\right\}}\left(1 - \exp{\left\{-\frac{N-1}{\alpha^\theta}\right\}}\right)\right],
\end{align*}
so $\sum_{\mathbf{K}\in\mathcal{K}_N}\sum_{\mathbf{J}\in\mathcal{J}_\infty - \mathcal{J}_M}\Pr_{\infty}(\mathbf{K},\mathbf{J}) + \sum_{\mathbf{K}\in\mathcal{K}_\infty-\mathcal{K}_N}\sum_{\mathbf{J}\in\mathcal{J}_\infty}\Pr_{\infty}(\mathbf{K},\mathbf{J})$, again by a similar argument as that used in the proof of Theorem 1, is also

\noindent $O\left(n\left(\exp{\left\{-\frac{N-1}{\alpha^\theta}\right\}} + \exp{\left\{-\frac{M-1}{\alpha^\psi}\right\}}\left(1 - \exp{\left\{-\frac{N-1}{\alpha^\theta}\right\}}\right)\right)\right)$.
\end{proof}

\begin{proof}[Details of Corollary 3]
The proof of Corollary 3 is identical to the proof of Corollary 2 through (11). In the remainder of the proof of Corollary 2, there are several steps in which we use an argument similar to that used in the proof of Theorem 1 to show that terms are \\
\noindent $O\left(n\left(\exp{\left\{-\frac{N-1}{\alpha^\theta}\right\}} + \exp{\left\{-\frac{M-1}{\alpha^\psi}\right\}}\left(1 - \exp{\left\{-\frac{N-1}{\alpha^\theta}\right\}}\right)\right)\right)$. At the corresponding steps of the proof of Corollary 3, we instead use arguments similar to that used in the proof of Corollary 1 to show that the corresponding terms are $O\left(n\left(\exp{\left\{-\frac{N-1}{\alpha^\theta}\right\}} + \exp{\left\{-\frac{M-1}{\alpha^\psi_{max}}\right\}}\left(1 - \exp{\left\{-\frac{N-1}{\alpha^\theta}\right\}}\right)\right)\right)$.
\end{proof}

\pagebreak

\section*{Appendix B. EDPM of Normals Truncation Approximation Posterior Sampling Computations}

In each iteration of the blocked Gibbs sampler, we sample:
\begin{enumerate}
    \item $(\boldsymbol{\mu}^*_y | \boldsymbol{\tau}^*_y, \mathbf{K}, \mathbf{Y})$: For each $k=1,\ldots,N$, draw
    \begin{align*}
        \boldsymbol{\beta}_{k}^* | \boldsymbol{\tau}_y^*, \mathbf{K}, \mathbf{Y} &\sim N\left(\left(\mathbf{x}_k^T\mathbf{x}_k + \mathbf{C}_y\right)^{-1}\left(\mathbf{x}_k^T\mathbf{y}_k + \mathbf{C}_y\boldsymbol{\beta}_0\right), \tau_{yk}^*\left(\mathbf{x}_k^T\mathbf{x}_k + \mathbf{C}_y\right)^{-1}\right),
    \end{align*}
    where $n_k$ is the number of observations currently assigned to the $k$th $\theta$-cluster.
    
    \item $(\boldsymbol{\tau}^*_y | \boldsymbol{\mu}^*_y, \mathbf{K}, \mathbf{Y})$: For each $k=1,\ldots,N$, draw
    \begin{align*}
        {\tau_{yk}^*}^{-1} | \boldsymbol{\mu}^*_y, \mathbf{K}, \mathbf{Y} &\sim Gamma\left( \nu_{y1} + \frac{n_k}{2}, \nu_{y2} + \sum_{i:K_i=k}\frac{(y_i-\mu_{yk}^*)^2}{2} \right).
    \end{align*}
    
    \item $(\boldsymbol{\mu}^*_{k} | \boldsymbol{\tau}^*_{xk}, \mathbf{K}, \mathbf{J}, \mathbf{X})$: For each $k=1,\ldots,N$, $j=1,\ldots,M$, and $l=1,\ldots,p$ draw
    \begin{align*}
        \mu_{j|k,l}^* | \boldsymbol{\tau}_{xk}^*, \mathbf{K}, \mathbf{J}, \mathbf{X} &\sim N\left(\left(\frac{n_{kj}}{\tau_{xj|k,l}^*} + \frac{1}{c_{x,l}}\right) \left(\frac{\sum_{i:K_i=k,J_i=j}x_{il}}{\tau_{xj|k,l}^*} + \frac{m_{l}}{c_{x,l}}\right), \left(\frac{n_{kj}}{\tau_{xj|k,l}^*} + \frac{1}{c_{x,l}}\right)^{-1}\right),
    \end{align*}
    where $n_{kj}$ is the number of observations currently assigned to the $j$th $\psi$-cluster within the $k$th $\theta$-cluster.
    
    \item $(\boldsymbol{\tau}^*_{xk} | \boldsymbol{\mu}^*_{xk}, \mathbf{K}, \mathbf{J}, \mathbf{X})$: For each $k=1,\ldots,N$ and $j=1,\ldots,M$, draw
    \begin{align*}
        {\tau_{xj|k,l}^*}^{-1} | \boldsymbol{\mu}^*_{xk}, \mathbf{K}, \mathbf{J}, \mathbf{X} &\sim Gamma\left( \nu_{x1} + \frac{n_{kj}}{2}, \nu_{x2} + \sum_{i:K_i=k,J_i=j}\frac{(x_{il}-\mu_{j|k,l}^*)^2}{2} \right).
    \end{align*}
    
    \item $(\mathbf{K}, \mathbf{J} | \mathbf{p}^\theta, \mathbf{p}^\psi, \boldsymbol{\mu}^*_y, \boldsymbol{\tau}^*_y, \boldsymbol{\mu}^*_{xk}, \boldsymbol{\tau}^*_{xk}, \mathbf{Y}, \mathbf{X})$: For each observation $i=1,\ldots,n$, draw a $\theta$- and $\psi$-cluster assignment, where the probability that observation $i$ is assigned to $\theta$-cluster $k$ and $\psi$-cluster $j$ is
    \begin{align*}
        \frac{p_k^\theta p_{j|k}^\psi}{\sqrt{\tau_{yk}^*\prod_{l=1}^p\tau_{xj|k,l}^*}}\exp{\left\{-\frac{1}{2}\left[ \frac{1}{\tau_{yk}^*}\left(y_i-\mathbf{x}_i\boldsymbol{\beta}_{k}^*\right)^2 + \sum_{l=1}^p\frac{1}{\tau_{xj|k,l}^*}\left(x_{il}-\mu_{j|k,l}^*\right)^2 \right] \right\}}.
    \end{align*}
    
    \item $(\mathbf{p}^\theta | \mathbf{K}, \alpha^\theta)$: To obtain draws for $\mathbf{p}^\theta$, we have $p_1^\theta=V_1^\theta$ and $p_k^\theta=V_k^\theta\prod_{h=1}^{k-1}(1-V_h^\theta)$, $k=2,\ldots,N$, where
    \begin{align*}
        V_k^\theta | \mathbf{K}, \alpha^\theta &\stackrel{ind}{\sim} Beta\left(n_k+1, \alpha^\theta + \sum_{h=k+1}^{N}n_h\right),
    \end{align*}
    $k=1,\ldots,N-1$, and $V_N^\theta=1$.
    
    \item $(\mathbf{p}^\psi_k | \mathbf{K}, \mathbf{J}, \alpha^{\psi|\theta}_k)$: For each $k=1,\ldots,N$, to obtain draws for $\mathbf{p}^\psi_k$, we have $p_{1|k}^\psi=V_{1|k}^\psi$ and $p_{j|k}^\psi=V_{j|k}^\psi\prod_{h=1}^{j-1}(1-V_{h|k}^\psi)$, $j=2,\ldots,M$, where
    \begin{align*}
        V_{j|k}^\psi | \mathbf{K}, \mathbf{J}, \alpha^\psi &\stackrel{ind}{\sim} Beta\left(n_{kj}+1, \alpha^{\psi|\theta}_k + \sum_{h=j+1}^{M}n_{kh}\right),
    \end{align*}
    $j=1,\ldots,M-1$, and $V_{M|k}^\psi=1$.
    
    \item $(\alpha^\theta|\mathbf{p}^\theta)$: Draw
    \begin{align*}
        \alpha^\theta | \mathbf{p}^\theta &\sim Gamma\left(N+\eta_{y1}-1, \eta_{y2}-\sum_{k=1}^{N-1}\log\left(1-V_k^\theta\right)\right).
    \end{align*}
    
    \item $(\alpha^{\psi|\theta}_k|\mathbf{p}^\psi)$ or $(\alpha^{\psi}|\mathbf{p}^\psi)$: If $\alpha^{\psi|\theta}_k$ depends on $\theta_k^*$, for each $k=1,\ldots,N$, draw
    \begin{align*}
        \alpha^{\psi|\theta}_k | \mathbf{p}_k^\psi &\sim Gamma\left( M+\eta_{x1}-1, \eta_{x2} - \sum_{j=1}^{M-1}\log\left(1 - V_{j|k}^\psi\right) \right),
    \end{align*}
    otherwise,
    \begin{align*}
        \alpha^{\psi} | \mathbf{p}^\psi &\sim Gamma\left( N(M-1)+\eta_{x1}-1, \eta_{x2} - \sum_{k=1}^N\sum_{j=1}^{M-1}\log\left(1 - V_{j|k}^\psi\right) \right).
    \end{align*}
\end{enumerate} 

\section*{Supplemental Table}
\captionsetup[table]{name=Supplemental Table}

\begin{table}[h]
    \centering
    \begin{tabular}{ccccccc}
         \toprule
          & \multicolumn{2}{c}{$p=5$} & \multicolumn{2}{c}{$p=10$} & \multicolumn{2}{c}{$p=15$} \\
         \cmidrule{2-7}
          & $\hat{l}_1$ & $\hat{l}_2$ & $\hat{l}_1$ & $\hat{l}_2$ & $\hat{l}_1$ & $\hat{l}_2$ \\
         \midrule
         BG sampler, $N=10$, $M=50$ & 0.004 & 0.001 & 0.006 & 0.005 & 0.012 & 0.010 \\
         BG sampler, varying $N$ and $M$ & 0.004 & 0.002 & 0.005 & 0.002 & 0.007 & 0.008 \\
         \bottomrule
    \end{tabular}
    \caption*{Supplemental Table 1: MC errors for estimates of prediction errors given in Table 2.}
    \label{tab:prederr}
\end{table}

\end{document}